\documentclass[fleqn,usenatbib]{mnras}

\usepackage[T1]{fontenc}

\DeclareRobustCommand{\VAN}[3]{#2}
\let\VANthebibliography\thebibliography
\def\thebibliography{\DeclareRobustCommand{\VAN}[3]{##3}\VANthebibliography}

\usepackage{graphicx}	% Including figure files
\usepackage{amsmath}	% Advanced maths commands
\usepackage{amssymb}	% Extra maths symbols
\usepackage{multirow}   % Splitting table rows
\usepackage{layouts}    % REMOVE??
\usepackage{comment}
\makeatletter
\newcommand\thefontsize[1]{{#1 The current font size is: \f@size pt\par}}
\makeatother

\DeclareTextFontCommand{\textttup}{\normalfont\ttfamily}
\makeatletter
\newcommand{\printexternalcurrentfont}{%
  \expandafter\format@externalcurrentfont\fontname\font:\@nil
}
\def\format@externalcurrentfont#1:#2\@nil{%
  \textttup{\@ifnextchar"{\@gobble}{}#1}%
}
\makeatother

\usepackage{newtxtext,newtxmath}

\newcommand{\afe}{\ifmmode{\left[\alpha/\rm{Fe}\right]}\else{$\left[\alpha/\rm{Fe}\right]$}\fi}

\newcommand{\dafe}{\ifmmode{\Delta\left[\alpha/\rm{Fe}\right]}\else{$\Delta\left[\alpha/\rm{Fe}\right]$}\fi}

\newcommand{\lre}{\ifmmode{\lambda_{R_{\rm{e}}}}\else{$\lambda_{R_{\rm{e}}}$}\fi}

\newcommand{\lreeo}{\ifmmode{\lambda_{\,R_{\rm{e}}}^{\rm{\,eo}}}\else{$\lambda_{\,R_{\rm{e}}}^{\rm{\,eo}}$}\fi}

\newcommand{\lreeofull}{\ifmmode{\lambda_{\,R_{\rm{e}}}^{\rm{\,edge-on}}}\else{$\lambda_{\,R_{\rm{e}}}^{\rm{\,edge-on}}$}\fi}

\defcitealias{Scott+17}{S17}
\defcitealias{McDermid+15}{M15}
\defcitealias{Bernardi+19}{B19}

\setcitestyle{notesep={ }}

\title[{\afe\ and \lre\ in the SGS}]{The SAMI Galaxy Survey: The Link Between \afe\ and Kinematic Morphology}

\author[P. J. Watson et al.]{Peter J. Watson,$^{1}$\thanks{E-mail: peter.watson2@physics.ox.ac.uk}
Roger L. Davies,$^{1}$
Jesse van de Sande,$^{2,4}$
Sarah Brough,$^{2,3}$
Scott M. Croom,$^{2,4}$
\newauthor Francesco D'Eugenio,$^{5,6}$
Karl Glazebrook,$^{2,7}$
Brent Groves,$^{8}$
\'Angel R. L\'opez-S\'anchez,$^{2,9,10}$
\newauthor Nicholas Scott,$^{2,4}$
Sam P. Vaughan,$^{4}$
C. Jakob Walcher,$^{11}$
Joss Bland-Hawthorn,$^{4}$
Julia J. Bryant,$^{2,4,12}$
\newauthor Michael Goodwin,$^{9}$
Jon S. Lawrence,$^{13}$
Nuria P. F. Lorente,$^{9}$
Matt S. Owers,$^{2,10,14}$
and Samuel Richards$^{15}$
\\
$^{1}$ Sub-Department of Astrophysics, Department of Physics, University of Oxford, Denys Wilkinson Building, Keble Rd., Oxford OX1 3RH, UK\\
$^{2}$ ARC Centre of Excellence for All Sky Astrophysics in 3 Dimensions (ASTRO 3D)\\
$^{3}$ School of Physics, University of New South Wales, NSW 2052, Australia\\
$^{4}$ Sydney Institute for Astronomy (SIfA), School of Physics, University of Sydney, NSW 2006, Australia\\
$^{5}$ Cavendish Laboratory and Kavli Institute for Cosmology, University of Cambridge, Madingley Rise, Cambridge, CB3 0HA, UK\\
$^{6}$ Sterrenkundig Observatorium, Universiteit Gent, Krijgslaan 281 S9, B-9000 Gent, Belgium\\
$^{7}$ Centre for Astrophysics \& Supercomputing, Swinburne University of Technology, Victoria 3122, Australia\\
$^{8}$ Research School of Astronomy \& Astrophysics, Australian National University, Mt Stromlo Observatory, Cotter Rd, Weston Creek, ACT 2611, Australia\\
$^{9}$ Australian Astronomical Optics, Macquarie University, 105 Delhi Rd, North Ryde, NSW 2113, Australia\\
$^{10}$ Department of Physics and Astronomy, Macquarie University, NSW 2109, Australia\\
$^{11}$ Leibniz-Institut f\"ur Astrophysik Potsdam (AIP), An der Sternwarte 16, D-14482 Potsdam, Germany\\
$^{12}$ Australian Astronomical Optics, AAO-USydney, School of Physics, Building A28, University of Sydney, NSW 2006, Australia\\
$^{13}$ Australian Astronomical Optics - Macquarie, Macquarie University, NSW 2109, Australia\\
$^{14}$ Astronomy, Astrophysics and Astrophotonics Research Centre, Macquarie University, Sydney, NSW 2109, Australia\\
$^{15}$ SOFIA Science Center, USRA, NASA Ames Research Center, Building N232, M/S 232-12, P.O. Box 1, Moffett Field, CA 94035-0001, USA\\
}

\date{Accepted XXX. Received YYY; in original form ZZZ}

\pubyear{2022}

\begin{document}
\label{firstpage}
\pagerange{\pageref{firstpage}--\pageref{lastpage}}
\maketitle

% Abstract of the paper
\begin{abstract}

We explore a sample of 1492 galaxies with measurements of the mean stellar population properties and the spin parameter proxy, \lre, drawn from the SAMI Galaxy Survey.
We fit a global \afe-$\sigma$ relation, finding that $\afe=(0.395\pm0.010)\rm{log}_{10}\left(\sigma\right)-(0.627\pm0.002)$.
We observe an anti-correlation between the residuals \dafe\ and the inclination-corrected \lreeo, which can be expressed as $\dafe=(-0.057\pm0.008)\lreeo+(0.020\pm0.003)$. 
The anti-correlation appears to be driven by star-forming galaxies, with a gradient of $\dafe\sim(-0.121\pm0.015)\lreeo$, although a weak relationship persists for the subsample of galaxies for which star formation has been quenched.
We take this to be confirmation that disk-dominated galaxies have an extended duration of star formation.
At a reference velocity dispersion of 200\,km\,s$^{-1}$, we estimate an increase in half-mass formation time from $\sim$0.5\,Gyr to $\sim$1.2\,Gyr from low- to high-\lreeo\ galaxies.
Slow rotators do not appear to fit these trends.
Their residual $\alpha$-enhancement is indistinguishable from other galaxies with $\lreeo\lessapprox0.4$, despite being both larger and more massive. 
This result shows that galaxies with $\lreeo\lessapprox0.4$ experience a similar range of star formation histories, despite their different physical structure and angular momentum.

\end{abstract}

% Select between one and six entries from the list of approved keywords.
% Don't make up new ones.

%%%%%%%%%%%%%%%%%%%%%%%%%%%%%%%%%%%%%%%%%%%%%%%%%%

%%%%%%%%%%%%%%%%% BODY OF PAPER %%%%%%%%%%%%%%%%%%

\section{Introduction}

\subsection{Background} \label{sec:background}

Understanding the formation history of individual galaxies requires both observations and modelling of their physical properties in three dimensions. Even in the era of large spectroscopic surveys we are limited by only being able to observe galaxies at a single point in their evolution.

One vital tool is that of stellar population analysis, which gives us an insight into the timescales and assembly history of the stellar component of galaxies.
Through comparison with semi-empirical and theoretical models, we can determine integrated properties such as the light-weighted age, metallicity, and elemental abundances.
Many models eschew the vast computational cost of calculating individual elemental abundances by grouping together elements with a similar formation mechanism.
Assuming that the quantity of these elements produced can be directly correlated to the number of supernovae of a given type, this significantly limits the number of free parameters.
The $\alpha$-element abundance is one of the most commonly utilised groupings.
These elements are predominantly formed in massive stars, prior to being ejected through core-collapse supernovae.
The \afe\ ratio compares the abundance of $\alpha$-elements to Fe, which is mainly formed over considerably longer timescales in Type Ia supernovae.
This abundance ratio therefore measures the relative contribution of each type of supernovae to the ISM, or more precisely, to the integrated light from stars that have formed from the ISM. 
Over a longer period of star formation, the relative contribution from Type Ia supernovae becomes more significant, and so \afe\ will tend towards solar values \citep{Greggio+83}.
For galaxies that are no longer actively forming stars, this measure therefore reflects the total duration of star formation prior to quenching \citep{de_la_Rosa+11}.

We can also determine the dynamical properties of a galaxy using stellar template libraries.
This particular field has a long and detailed history, dating back to the first discovery of ``nebular rotation'' \citep{Slipher+14}.
More recently, with the advent of integral field spectroscopy, the SAURON \citep{de_Zeeuw+02} and ATLAS$^{\rm{3D}}$ \citep{Cappellari+11} surveys enabled a quantitative classification of galaxies based on their kinematics.
Based on features in the velocity field \citep{Krajnovic+11}, galaxies were separated into fast rotators, with ordered rotation and disks, and slow rotators, with more complex velocity fields \citep{Emsellem+11}.
The selection criteria have since been refined \citep{Cappellari+16,van_de_Sande+21a}, but the presence of a distinct bimodality in the kinematic distribution of galaxies has remained \citep{Graham+18}, suggesting multiple formation scenarios.

There have been many studies whose motivation has been to investigate the build-up of mass and angular momentum, in order to better understand the origin of present-day morphologies \citep[][contains a thorough review]{Naab+14}.
There are, however, fewer studies linking galaxy kinematics and elemental abundance ratios.
\cite{Eggen+62} investigated the motion of stars within the Milky Way, finding an anti-correlation between the ellipticity of an orbit and the inferred stellar metallicity.
They utilised this result in a model galaxy to estimate the formation timescale, providing one of the first links between stellar dynamics and stellar properties.

\cite{Franx+90} made the assumption that colour differences within and between galaxies were due only to changes in the stellar metallicity, and found a correlation between this inferred metallicity, and the local escape velocity.
This was confirmed by \cite{Davies+93}, using a spectroscopic measurement of metallicity, alongside the observation that the line indices Mg$_2$ and $\langle\rm{Fe}\rangle$ coupled differently to the galaxy kinematics.
\cite{Trager+00} followed on from this with their findings of a simple scaling relation between the stellar velocity dispersion $\sigma$ of early-type galaxies (ETGs), and the abundance ratio \afe.
More recently, studies such as \citet[][hereafter:\ \citetalias{McDermid+15}]{McDermid+15} have focussed on the internal kinematics of early-type galaxies, and have shown that non-regularly rotating galaxies are offset to lower metallicities compared to the global mass-metallicity relation.

Here, we focus primarily on the \afe-$\sigma$ relation.
Since its initial discovery, it has proven to be a useful test for simulations and semi-analytic models of galaxy formation, despite some difficulties in reproducing this relation across different mass scales \citep{Segers+16}.
Whilst comparisons across size, optical morphology, and environment are well researched \citep[\textit{e.g.}][]{Annibali+11,Scott+17,Sanchez+21,Watson+21}, the $\alpha$-enhancement as a function of galactic dynamics has not been investigated as thoroughly.

In \citetalias{McDermid+15}, the authors also looked at offsets from a global \afe-$\sigma$ relation, using the regular/non-regular rotator classification from \cite{Krajnovic+11}.
Here, using a sample of 260 ETGs, they found no measurable difference between the two kinematic classes.
\citet[][hereafter:\ \citetalias{Bernardi+19}]{Bernardi+19}, after stacking spectra from elliptical galaxies in the SDSS-IV MaNGA survey \citep{Bundy+15}, instead found slow rotators were $\alpha$-enhanced by 0.04\,dex.
However, they also used a different selection criteria to \cite{Cappellari+16}, in order to prevent contamination of their fast rotator sample, which may have the unintended effect of obscuring or altering any underlying trends.
\cite{Krajnovic+20} briefly touched on the \afe-$\sigma$ relation, but showed no obvious difference in the residuals \dafe\ between fast and slow rotators.

Perhaps the greatest limitation on these previous studies has been the morphological selection, with many only investigating ETGs.
Considering later morphological types, we note that substantial number of spiral and lenticular galaxies can be separated out into two distinct components, a central bulge, and an extended disk.
Under the inside-out scenario of galaxy formation, so-called classical bulges are thought to form through violent gas collapse or mergers \citep{Larson+74,Bender+92} over short timescales, with the disk gradually building up around them.
In this scenario, disk-dominated galaxies will have had a longer overall duration of star formation relative to bulge-dominated galaxies, and show a greater degree of rotational support.
we can therefore use \afe\ as an indicator of the duration of star formation, and \lre\ as a measure of the rotational support, to explore the extent to which this formation scenario is supported by the observational evidence.

In order to determine the kinematic classification, we require the full spatial information afforded by integral-field spectroscopy (IFS).
There exist several large IFS surveys to date, including the CALIFA Survey \citep{Sanchez+12}, ATLAS$^{\text{3D}}$ \citep{Cappellari+11}, and SDSS-IV MaNGA \citep{Bundy+15}.
Here, we present results drawn from the SAMI (Sydney-AAO Multi-object Integral field spectrograph) Galaxy Survey \citep[SGS,][]{Bryant+15}. 
In Section \ref{sec:Data} we describe the SGS in more detail, including the sources of the ancillary data used.
The methods used for extracting both the luminosity-weighted stellar population parameters and the kinematics are summarised in Section \ref{sec:method}.
We present our results, the dependence of \afe\ on the dynamical properties of galaxies in \ref{sec:kinematic_morphologies}.
Finally, we discuss the implications of our research and conclude in Sections \ref{sec:discussion} and \ref{sec:conclusions}.
In line with other SAMI team papers, we assume a $\Lambda$CDM cosmology, with $\Omega_{\rm{m}}=0.3$, $\Omega_{\Lambda}=0.7$, and $H_0=70$\,km\,s$^{-1}$Mpc$^{-1}$.
All magnitudes given are in the AB system \citep{Oke+83}, and stellar masses and star-formation rates assume the initial mass function of \cite{Chabrier+03}.

\section{Data}
\label{sec:Data}

\subsection{SAMI Galaxy Survey} 
\label{sec:SAMI_galaxy_survey}

The SAMI instrument and survey design are detailed extensively in both \cite{Croom+12} and \cite{Bryant+15}.
The instrument comprises 13 Integral Field Units (IFUs, also known as \textit{hexabundles} in SAMI), which can be deployed over a 1-degree diameter field of view, each with an individual field of view of 15 arcsec \citep{Bland_Hawthorn+11, Bryant+14}.
The IFUs are mounted at the prime focus of the Anglo-Australian Telescope (AAT), and each consists of 61 individual fibres.
Observations are dithered to create data cubes with a 0.5-arcsec spaxel size.
All 819 fibres (including 26 allocated to blank sky observations for calibration purposes) are fed into the AAOmega spectrograph \citep{Saunders+04, Smith+04, Sharp+06}.
This is composed of a blue arm, with spectral resolution $R\sim1800$ over the wavelength range 3750-5750\,\AA, and a higher resolution red arm, with wavelength coverage 6300-7400\,\AA\ and $R\sim4300$ \citep{van_de_Sande+17b}.

The SGS consists of 3426 observations of 3068 unique galaxies, available as part of the SAMI public data releases \citep{Allen+15, Green+18, Scott+18, Croom+21}.
The survey spans a redshift range $0.004<z<0.115$, and a stellar mass range $M_*\sim10^7$ to $10^{12}\,M_{\odot}$.
Field and group galaxies were drawn from the Galaxy And Mass Assembly (GAMA) survey \citep{Driver+11}, with the selection being volume-limited in each of four stellar mass cuts.
An additional sample of cluster galaxies was drawn from the survey of eight low-redshift clusters in \cite{Owers+17}, to extend the environmental sampling.

\subsection{Ancillary data}
\label{subsec:ancillary_data}

We make use of additional measurements by other members of the SAMI team throughout our analysis.
These include circularised effective radii ($r_e$), measured using the Multi Gaussian Expansion \citep[MGE,][]{Emsellem+94} algorithm of \cite{Cappellari+02} and photometric fits by \cite{D'Eugenio+21} to $r$-band images from either SDSS Data Release 9 \citep{Ahn+12} or VST ATLAS surveys \citep{Shanks+13, Shanks+15}.
Optical morphological classifications are taken from SAMI Public Data Release 3 \citep{Croom+21}, following the method of \cite{Cortese+16}, where galaxies were designated as one of four types (Ellipticals, S0s, early- and late-type spirals).
This was based on visual inspection of colour images by $\sim$10 SAMI team members, taken from the same sources as the photometric fits.
Galaxies were assigned an integer between 0 (for ellipticals) and 3 (for late-type spirals and irregulars), with half-integers reserved for galaxies where the classification was split between two morphological types.

Stellar masses are taken from the SGS sample catalogue \citep{Bryant+15}.
These were derived from the rest-frame \textit{i}-band absolute magnitude and $g-i$ colour by using the colour-mass relation following the method of \cite{Taylor+11}.
Measurements of the global star-formation rate are taken from DR3, using the method described in \cite{Medling+18}, and are based on extinction-corrected H$\alpha$ fluxes, which are converted to SFRs using the relation of \cite{Kennicutt+94}.

\section{Method}
\label{sec:method}

\subsection{Stellar populations}
\label{subsec:stell_pops}

Stellar population measurements are taken from \cite{Watson+21}, using an approach based on measurements of absorption line indices.
We briefly summarise the method here.

We utilise 20 Lick indices defined by \cite{Worthey+97} and \cite{Trager+98}, which fall within the SAMI wavelength range.
This comprises five Balmer lines (H$\delta_A$, H$\delta_F$, H$\gamma_A$, H$\gamma_F$, H$\beta$), six iron-dominated indices (Fe4383, Fe4531, Fe5015, Fe5270, Fe5335, Fe5406), and the molecular and elemental lines CN1, CN2, Ca4227, G4300, Ca4455, C4668, Mg1, Mg2, and Mg$_b$.
The galaxies were corrected for emission-line infill and bad pixels through a three-fold fit using \textsc{pPXF} \citep{Cappellari+17}, and the MILES empirical stellar spectra of \cite{Vazdekis+11}.
The initial fit weighted all pixels equally, and the standard deviation of the residuals allowed us to determine a scaling factor for the noise spectrum.
The second fit weighted pixels according to this noise spectrum, and employed a 3-sigma clipping method to determine bad pixels on the detector, and those contaminated by emission-line infill.
For the final fit, these regions were masked out, and the resulting best-fit output from \textsc{ppxf} was used to replace the missing values.
This method of replacing anomalous pixels, based on emission-free template spectra, is more robust than interpolation, or subtracting fitted emission lines, particularly for low S/N spectra.

The cleaned galaxy spectra were then broadened to the required Lick/IDS resolution \citep{Worthey+97}, by convolution with a wavelength-dependent Gaussian.
This convolution accounted for the instrumental resolution, and intrinsic broadening due to velocity dispersion.
For galaxies and indices where the combined broadening already exceeded the Lick/IDS resolution, a correction factor was applied \citep[see][for further details]{Watson+21}.
The indices were measured using the variance-weighted method of \cite{Cenarro+01}.
The errors on the indices were estimated following a bootstrap procedure, in which noise was randomly added to the galaxy spectra, and the indices remeasured.
After 100 realisations, the standard deviation was taken as the error on each index.

The indices were used to predict simple stellar population (SSP) parameters using the models of \cite{Thomas+10}.
These models predict Lick index measurements as a function of $\rm{log}_{10}\left(\rm{age}\right)$, $\left[\rm{Z/H}\right]$, and \afe.
The $\chi^2$ minimisation procedure of \cite{Proctor+04} was utilised to find the best-fitting combination of SSP parameters.
Galaxies were rejected if fewer than five indices were available, or if the indices did not
include at least one Balmer index, and one Fe index.
Indices were also rejected from the fit if they lay more than 1$\sigma$ outside the model grid, so as not to bias the solution.

\subsection{Kinematics}
\label{subsec:method_kinematics}

Measurements of the stellar velocity dispersion, $\sigma$, are taken from SAMI DR3 \citep{Croom+21}, and are measured from spectra within an $r_e$ aperture.
These values are measured following an almost identical method to that in Section \ref{subsec:stell_pops}, which is described in detail in \cite{van_de_Sande+17b}.
The primary difference extends from the treatment of the emission lines, which are masked out entirely for measuring the stellar kinematics.
Errors are estimated following a Monte-Carlo approach.
The spectrum is divided into 14 regions, with residuals from the final fit randomly reallocated within each region.
These residuals are added to the original galaxy spectrum, then $\sigma$ can be remeasured using the best fit template.
This process is repeated 100 times, with the standard deviation of the new measurements taken as the error on $\sigma$.

Measurements of the spin parameter proxy, \lre, are also taken from SAMI DR3, again using a method from \cite{van_de_Sande+17b}.
These values are derived using the definition of \cite{Emsellem+07}:
\begin{equation}
    \lambda_R = \frac{ \langle R |V| \rangle }{ \langle R \sqrt{V^2 + \sigma^2 } \rangle}
    = \frac{ \sum_{i=0}^{N_{\rm{spx}}} F_i R_i |V_i| }{ \sum_{i=0}^{N_{\rm{spx}}} F_i R_i \sqrt{V_i^2 + \sigma_i^2 } },
\end{equation}
where $i$ is the spaxel position, $F_i$ is the flux of the $i^{\rm{th}}$ spaxel, and $V_i$ and $\sigma_i$ are the stellar velocity and velocity dispersion in km\,s$^{-1}$.
The definition of $R_i$ is adopted from \cite{Cortese+16}, in that it is the semimajor axis of the ellipse on which each spaxel sits, as opposed to the circular projected radius as originally used by \cite{Emsellem+07}.
\lre is calculated using an elliptical aperture with semimajor axis $R_{\rm{e}}$, with all spaxels that pass the kinematic quality cuts $Q_1$ and $Q_2$ \citep{van_de_Sande+17b}.
A fill factor of 95\% of good spaxels within the aperture is required for these measurements.
When the largest available radius is smaller than the effective radius $R_{\rm{e}}$, an aperture correction is applied, following the method described in \cite{van_de_Sande+17a}.
The measurements we use have also been corrected for seeing, with the method described in \cite{Harborne+20a}.

We also make considerable use of \lreeofull, the \lre measurements corrected to an edge-on projection, and hereafter referred to as \lreeo. 
These values are obtained following the method described in \cite{van_de_Sande+18} and \cite{van_de_Sande+21b}.
They are derived from the observed \lre and $\epsilon_e$ measurements, assuming theoretical model predictions for galaxies as rotating, oblate spheroids with varying intrinsic shape and anisotropy. 
No correction is applied (\textit{i.e.} $\lreeo=\lre$) for galaxies that are not consistent with being simple rotating spheroids.

\subsection{Line fitting} 
\label{subsec:line_fitting}

For all linear fits, we make use of the Python library \textsc{lmfit} by \cite{lmfit}, where we minimise the quantity
\begin{equation} \label{eq:lmfit_chi}
    \chi^2 = \sum^N_{j=1} \frac{\left[a(x_j-x_0)+b-y_j\right]^2}{
    (a\Delta x_j)^2 + (\Delta y_j)^2},
\end{equation}
adopted from \cite{Tremaine+02}, which accounts for errors in both $x$ and $y$.
We set $x_0$ to zero unless otherwise stated, to simplify comparisons throughout the paper and with other studies.
The inclusion of additional terms for intrinsic scatter, or minimising the scatter orthogonal to the relation, does not change our results.

\subsection{Completeness} \label{subsec:completeness}

\begin{table*}
    \centering
    \caption[]{
    The total number of galaxies used for each stage of the analysis, separated by both optical and kinematic morphology. 
    Intermediate classifications have been grouped with the earlier of the two types, e.g. E/S0 galaxies contribute to the total under the E column.
    The selection criteria for the kinematic classifications are also shown.
    These divisions were chosen to give a similar spread of \lreeo across each Region.
    Note that Region I does not include galaxies already classified as slow rotators.
    }
    \label{tab:total_number_of_galaxies}
    \begin{tabular}{rccccccl}\hline\hline
        \multirow{2}{*}{Sample} & \multirow{2}{*}{Total} & \multicolumn{5}{c}{Optical Morphology} & \multirow{2}{*}{Section} \\\cline{3-7}
        & & E & S0 & Sa/b & Sc & Unclassified &\\\hline
        Unique Galaxies & 3068 & 561 & 728 & 605 & 1026 & 148 &\\
        Kinematic Sample & 1492 & 323 & 480 & 467 & 191 & 31 & \ref{sec:kinematic_classification} \\
        \multirow{2}{*}{} & \multirow{3}{*}{} & \multicolumn{5}{c}{Kinematic Morphology} &\\\cline{3-7}
        & & SR & I & II & III & IV &\\
        \lreeo\ Range & & Eq.~\ref{eq:SR_selection} & $\leq0.4$ & 0.4-0.525 & 0.525-0.65 & $>0.65$ &\\\cline{3-7}
        Kinematic Sample & 1492 & 121 & 218 & 302 & 432 & 419 & \ref{sec:kinematic_classification} \\
        Quenched Sample & 657 & 88 & 125 & 196 & 175 & 73 & \ref{sec:quenched_galaxies} \\\hline
    \end{tabular}
\end{table*}

\begin{figure}
    \centering
    \includegraphics[width=\columnwidth]{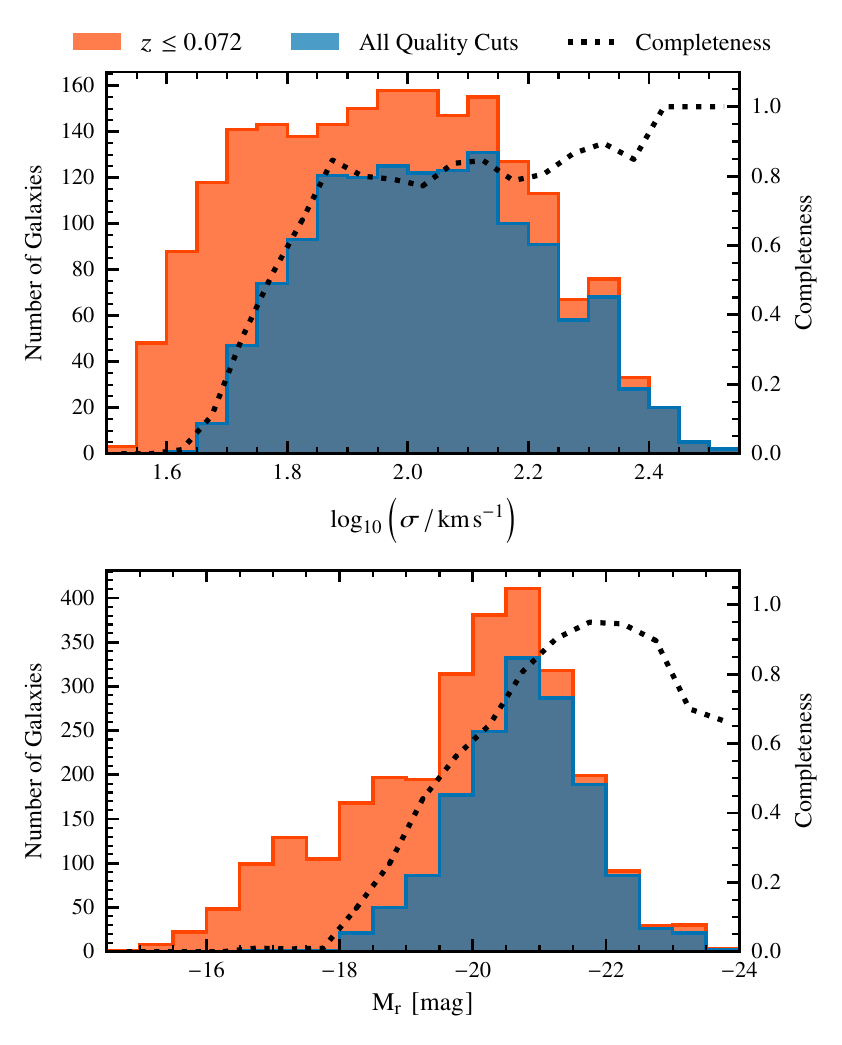}
    \caption[]{The number of unique galaxies in both the redshift-limited sample, and the core sample after all quality cuts, and their distribution as a function of $\rm{log}_{10}\left(\sigma\right)$, and absolute magnitude in the $r$-band, $M_r$.
    The dotted line indicates the completeness of our final sample after all quality cuts, relative to the number of unique galaxies with $z\leq0.072$.}
    \label{fig:quality_cut_hist}
\end{figure}

We make a series of cuts to the overall sample of 3068 unique galaxies in the SAMI Galaxy Survey.
In \cite{Watson+21}, we established that galaxies at redshifts $z>0.072$ showed strong signs of skyline contamination, which affected our measurements of \afe.
We therefore reject these galaxies, leaving 2773 galaxies with $z\leq0.072$, which we take as our baseline for completeness.

We apply the recommended quality cuts from \cite{van_de_Sande+17b} to the kinematic measurements.
These include $Q_1$, which requires the line-of-sight velocity $v$ to have a maximum uncertainty $v_{\rm{error}}<30$\,km\,s$^{-1}$, and $Q_2$, that the uncertainty on the velocity dispersion $\sigma$ follows $\sigma_{\rm{error}}<0.1\sigma_{\rm{obs}}+25$\,km\,s$^{-1}$.
The cumulative effect of $Q_1$ and $Q_2$ reduces the sample size to 1566 galaxies.

We also apply the quality cuts from \cite{Watson+21}.
We remove 7 galaxies with a spectral $\rm{S/N}<20$, and 1 galaxy with $\rm{log}_{10}\left(\sigma\right)<1.6$. 
We further remove 20 galaxies which either fall on the outer limits of the parameter space in \afe\ or have uncertainties spanning the entire range.
This leaves us with 1538 galaxies, with the cumulative effect shown in Figure \ref{fig:quality_cut_hist}.
We measure the completeness against the 2773 galaxies with $z\leq0.072$. 
The culled sample is $>80$\% complete for $\sigma>70$\,km\,s$^{-1}$, and for $M_r<-21$.
The completeness of the sample suffers at low $\sigma$, and similarly for the magnitude distribution, we find $>90\%$ of galaxies fainter than $M_r=-18$ are cut.

Since our analysis depends on fitting a reliable relationship between \afe\ and $\sigma$, we enforce a final cut to our sample.
We reject all galaxies with $\rm{log}_{10}\left(\sigma\right)<1.75$, such that we have a completeness of >50\% throughout the full range of velocity dispersion.
The remaining 1492 galaxies are referred to as the `kinematic sample'.
Comparing against the sample of 2093 galaxies used in \cite{Watson+21}, we have cut a substantially higher number of galaxies at low velocity dispersion due to the additional constraints from \cite{van_de_Sande+17b}.
The number of galaxies of each type are summarised in Table \ref{tab:total_number_of_galaxies}, where we can see the effect of the quality cuts on the sample morphology distribution.
Galaxies classified as Sc are largely removed from the sample, because they typically have $\rm{log}_{10}\left(\sigma\right)<1.75$.

\section{Results} \label{sec:kinematic_morphologies}

\subsection{Kinematic classification} \label{sec:kinematic_classification}

\begin{figure}
    \centering
    \includegraphics[width=\columnwidth]{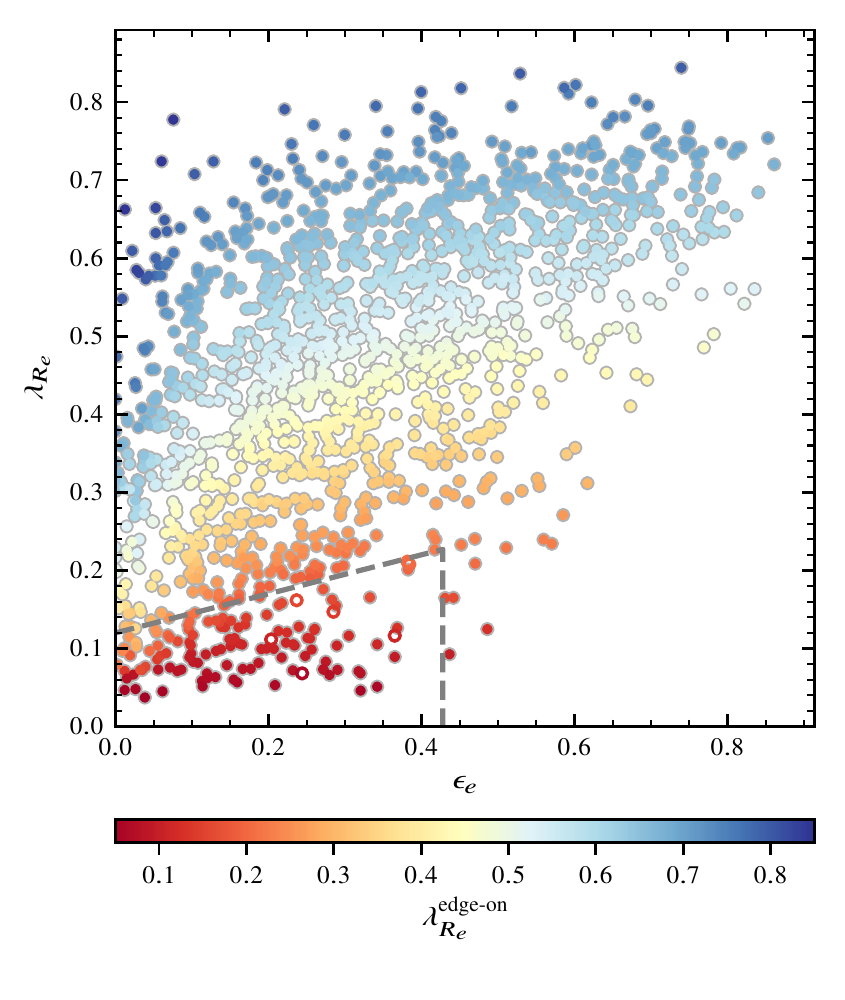}
    \caption[]{All galaxies in the SGS, for which \lre\ measurements exist, are presented here in the \lre-\,ellipticity plane. 
    Galaxies are coloured according to their \lreeo\ values.
    The dashed line delineates the slow rotator classification from \cite{van_de_Sande+21a}.
    Open circles represent galaxies which were revealed to be face-on spirals in a visual inspection, and as such are not included in our analysis.}
    \label{fig:SR_selection}
\end{figure}

We first use the selection criteria from \cite{van_de_Sande+21a}, which have been optimised for use with SGS data.
We classify galaxies as Slow Rotators (SRs) if they lie in the region of the \lre-\,ellipticity plane delineated by 
\begin{equation} \label{eq:SR_selection}
    \lre<\lambda_{R_{\text{start}}}+\epsilon_e/4,\ \text{with}\ \epsilon_e<0.35+\frac{\lambda_{R_{\text{start}}}}{1.538},
\end{equation}
where $\lambda_{R_{\text{start}}}=0.12$.
We show this in Fig.~\ref{fig:SR_selection}, noting that our results do not substantially change if we use instead the selection criteria of \cite{Cappellari+16}.
The remaining galaxies are typically classified as Fast Rotators (FRs).
For our analysis, rather than treating the FRs as a monolithic block, we further subdivide them into four separate groups, sorted by \lreeo and denoted as Region I-IV, with the selection criteria displayed in Table \ref{tab:total_number_of_galaxies}.
These criteria were chosen such that a similar range of \lreeo\ was surveyed in each group.

\subsection{Global relations} \label{sec:global_relations}

\begin{figure}
    \centering
    \includegraphics[width=\columnwidth]{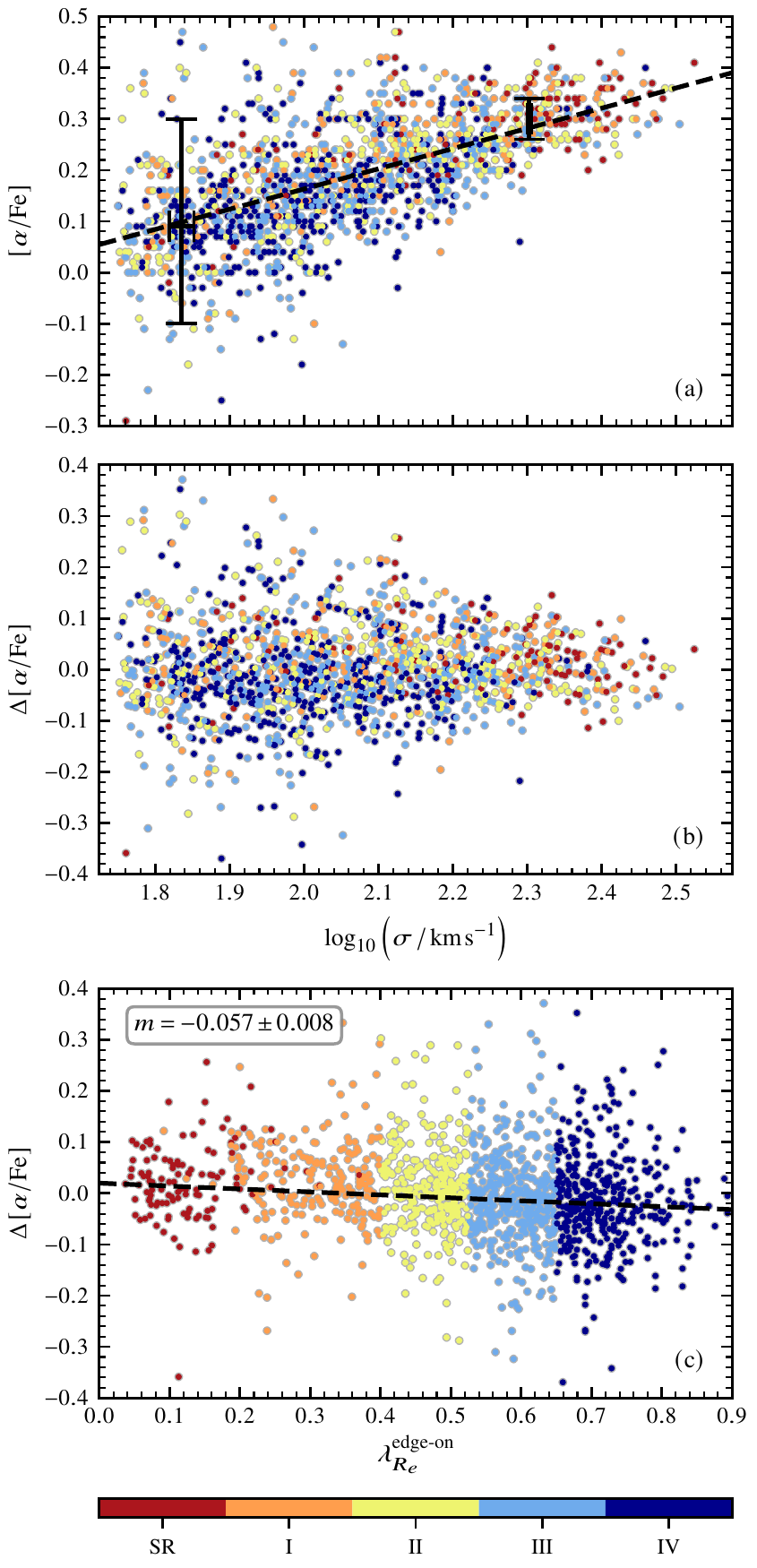}
    \caption[]{
    Galaxies are coloured according to their \lreeo\ values, using the region classification scheme from Table \ref{tab:total_number_of_galaxies}, with slow rotators (SR) marked in red, and Region IV galaxies in dark blue. 
    The colour scheme is consistent across all four plots.
    (a) The $\sigma$-\afe\ relation for our kinematic sample, with the best fit linear relation represented by the dashed line.
    (b) The residuals \dafe, as a function of $\text{log}_{10}\left(\sigma\right)$.
    There is no correlation, as expected, although the scatter increases at lower values of $\sigma$ due to the decrease in $S/N$.
    (c) The residuals as a function of \lreeo. 
    The best fit relation is shown as the dashed line, with the gradient $m$ inset.
    }
    \label{fig:sigma_lambda_r_residuals_base}
\end{figure}

We fit a weighted linear relationship between \afe\ and $\rm{log}_{10}(\sigma)$ to all 1,492 galaxies in our final kinematic sample, resulting in  $\afe=(0.395\pm0.010)\rm{log}_{10}\left(\sigma\right)-(0.627\pm0.002)$.
We show the result in Fig.~\ref{fig:sigma_lambda_r_residuals_base}, alongside the residuals \dafe\ as a function of $\sigma$.

Observing the distribution of the residuals overall, we can see that the scatter decreases with increasing $\sigma$, as $\sigma$ and the spectral S/N are positively correlated across the SGS.
Otherwise, there is no underlying structure evident, and the residuals are evenly distributed around 0, as we would expect from the model underlying LMFIT.

In Fig.~\ref{fig:sigma_lambda_r_residuals_base}, galaxies are coloured according to their measurements of the inclination-corrected spin proxy \lreeo, using the region classification scheme from Table \ref{tab:total_number_of_galaxies}.
From this, we can see that there is a clear dependence on \lreeo in the residuals in Fig.~\ref{fig:sigma_lambda_r_residuals_base}b, with Region III and IV galaxies predominantly having $\dafe\lessapprox0$.
Conversely, although less pronounced, Region I galaxies are weighted towards higher values of \dafe.

Therefore, if we instead present the residuals as a function of \lreeo, as in Fig.~\ref{fig:sigma_lambda_r_residuals_base}c, we find a statistically-significant anti-correlation.
Fitting a linear trendline to this gives the result $\dafe=(-0.057\pm0.008)\lreeo+(0.020\pm0.004)$.
The scatter around this relationship increases slightly towards higher values of \lreeo, although this is not as pronounced as the effect seen in Fig.~\ref{fig:sigma_lambda_r_residuals_base}b.
If we use the projected values of \lre\ rather than the inclination-corrected \lreeo, the gradient is $\sim50$\% steeper.
From Fig.~\ref{fig:SR_selection}, this is the expected outcome of applying the correction, since many galaxies have been shifted to higher numerical values of \lreeo.

\subsection{Slow rotators} \label{subsec:SRs}

\begin{figure}
    \centering
    \includegraphics[width=\columnwidth]{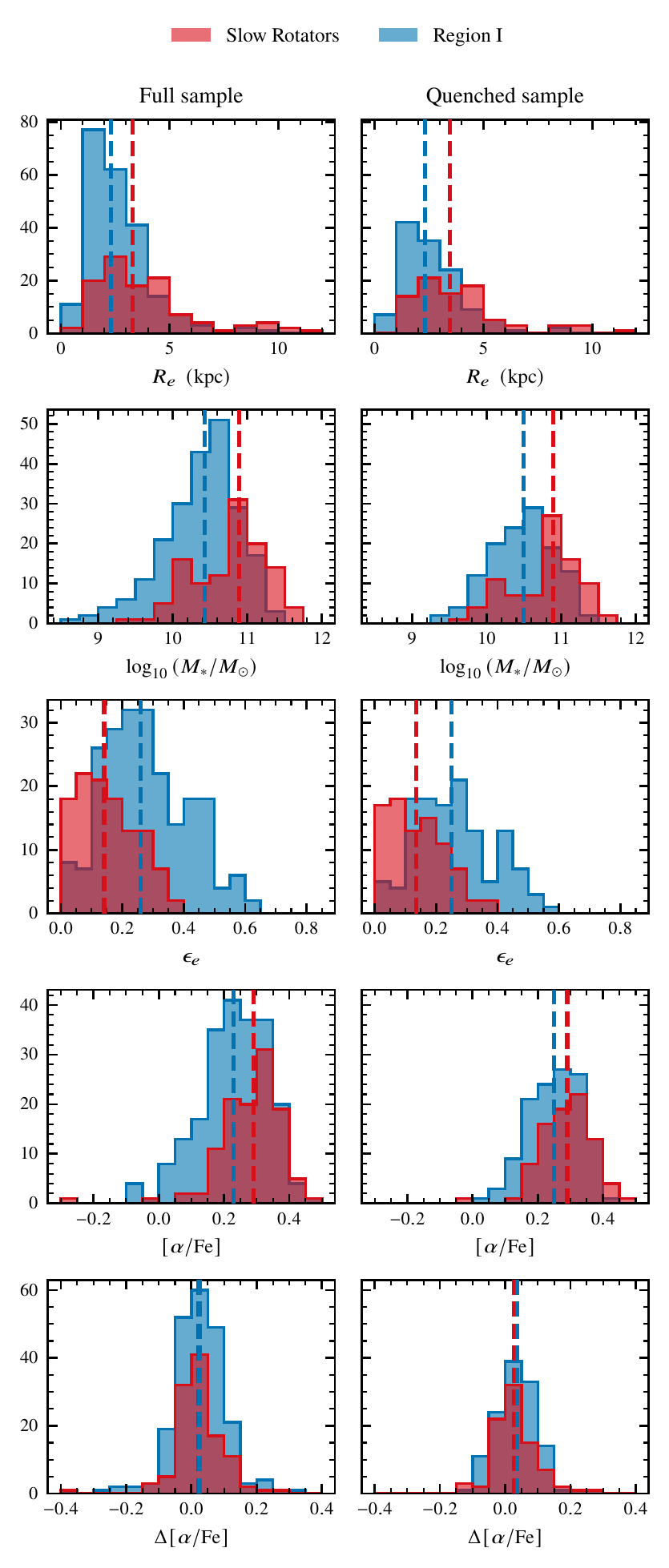}
    \caption[]{
    On the left hand side, we show the difference in the distributions of $R_e$, $\rm{log}_{10}(M_*/M_\odot)$, $\epsilon_e$, \afe, and \dafe\ for Slow Rotators and Region I galaxies in the full kinematic sample.
    The dashed lines indicate the medians of each distribution.
    On the right, we show the same properties after we have isolated the quenched population (see Section \ref{sec:quenched_galaxies}), illustrating that the two populations still show distinct distributions of physical parameters, but no measurable offset in the residual $\alpha$-enhancement, \dafe.
    }
    \label{fig:slow_reg_I_hist_comp}
\end{figure}

We also consider the effect of including slow rotators in this relationship, as they are kinematically distinct from the fast rotators in the remainder of the sample.
We begin by a comparison against galaxies denoted as ``Region I'', as discussed in Section \ref{sec:kinematic_classification}.
We show the mass and effective radii distributions of the two groups in Fig.~\ref{fig:slow_reg_I_hist_comp}.
Region I galaxies have a median effective radius of 2.3\,kpc, compared to SRs with a median radius of 3.3\,kpc.
SRs are much rounder than Region I galaxies, with median ellipticities of 0.14 and 0.26 respectively.
Similarly, SRs have higher stellar masses, showing a much more sharply peaked distribution with a median stellar mass of $10^{10.9}\,M_\odot$, compared to $10^{10.4}\,M_\odot$ for Region I.
This distribution of mass complicates the comparison of \afe\ across the two groups. 
Since we already know that galaxies with higher stellar mass, or velocity dispersion, have a higher $\alpha$-enhancement on average, we expect the median \afe\ of the SRs to be higher.
The measured difference of 0.06\,dex, whilst significant, is therefore not unexpected.

\begin{table}
    \centering
    \caption{The median values of \lreeo, \afe\ and \dafe\ for each subsample selected in Section \ref{sec:kinematic_classification}, where the residuals are measured from the global relation shown in Fig.~\ref{fig:sigma_lambda_r_residuals_base}.}
    \label{tab:med_values}
    \begin{tabular}{rccc}\hline\hline%|r|Y|Y|}\hline
        Group & \lreeo &$\afe_{\rm{med}}$ & $\dafe_{\rm{med}}$\\\hline
        Slow Rotators &0.105& 0.29 &  0.022\\
        Region I & 0.319 & 0.23 & 0.025\\
        Region II & 0.483 & 0.22 & 0.003 \\
        Region III & 0.591 & 0.19 & -0.010\\
        Region IV & 0.712 & 0.14 & -0.024 \\\hline
    \end{tabular}
\end{table}

To make a fair comparison, we instead look at the distribution of the residuals \dafe, from the global $\afe$-$\sigma$ relation displayed in Fig.~\ref{fig:sigma_lambda_r_residuals_base}.
The distributions are shown in Fig.~\ref{fig:slow_reg_I_hist_comp}, and the medians for all regions are given in Table \ref{tab:med_values}.
We notice that despite having a higher $\afe_{\rm{med}}$, the median residual for SRs is almost identical to that of Region I galaxies.
Thus, in the region of parameter space where the distributions of galaxy velocity dispersion $\sigma$ overlap, there is no measurable difference between the \afe\ ratio of SRs and Region I galaxies, in contrast to the relation seen across the fast rotators.
We therefore infer that at low \lreeo\ ($\lreeo<0.4$), any further separation of galaxies by their degree of rotational support, i.e. Region I galaxies and slow rotators, has no bearing on the \afe\ ratio (beyond that arising from the correlation with velocity dispersion), and hence the most probable evolutionary pathway of the stellar component.

The mass and radii distributions for Region I galaxies, shown in Fig.~\ref{fig:slow_reg_I_hist_comp}, are also consistent with galaxies in Regions II-IV.
The median $\alpha$-enhancement and residual \afe\ are given in Table \ref{tab:med_values}.
As we move from $\lreeo|_{\rm{med}}\sim0.3$ to higher values, we find that $\dafe_{\rm{med}}$ decreases monotonically.
Taken on its own, this would seem to imply a link between the duration of star formation and the angular momentum of galaxies in Regions I-IV.

\subsection{Relations for the quenched sample} \label{sec:quenched_galaxies}

\begin{figure}
    \centering
    \includegraphics[width=\columnwidth]{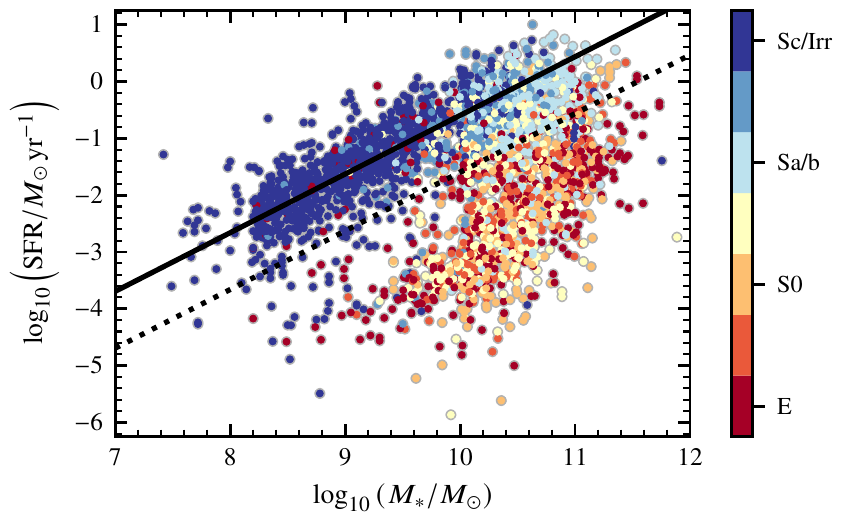}
    \caption{
    All galaxies in the SGS shown in the SFR-$M_*$ plane. 
    We colour the galaxies according to their optical morphology, and fit the SFMS (denoted as the solid black line) to those classified as Sc/Irr.
    The dotted black line shows the 1\,dex cutoff in SFR, below which we classify galaxies as quenched.}
    \label{fig:sfms_linefit_sb_col_vis_morph}
\end{figure}

\begin{figure}
    \centering
    \includegraphics[width=\columnwidth]{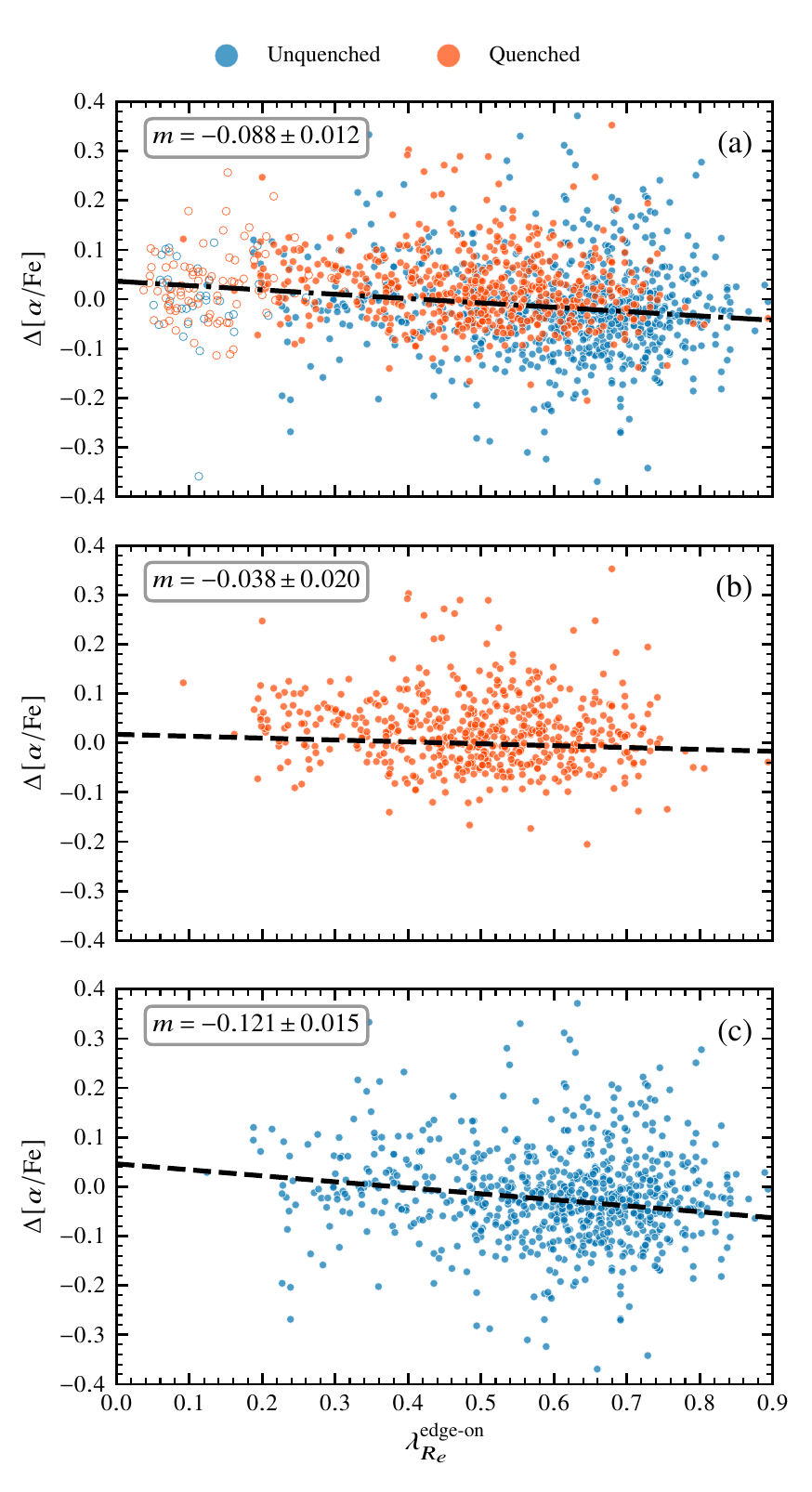}
    \caption[]{
    (a) The correlation between the residuals \dafe\ and \lreeo, using the global \afe-$\sigma$ relation from Fig.~\ref{fig:sigma_lambda_r_residuals_base}a.
    The dot-dashed line shows the best-fit linear relation, with the gradient $m$ inset.
    Following the discussion in Section \ref{subsec:SRs}, slow rotators are not included in the derivation of this relation, but are displayed here as open circles for reference.
    The galaxies shown are therefore identical to those in  Fig.~\ref{fig:sigma_lambda_r_residuals_base}c, with a different linear relation.
    Galaxies are coloured according to their classification as ``quenched'' or ``unquenched''.
    (b, c) As above, but the linear trendlines are fitted only to the quenched and unquenched galaxies respectively.
    }
    \label{fig:quenched_residuals}
\end{figure}

We isolate a quenched subsample of the SGS, utilising the measurements of the star-formation rate (SFR) from SAMI DR3.
We display all SGS galaxies in the SFR-M$_*$ plane in Fig.~\ref{fig:sfms_linefit_sb_col_vis_morph}, where we have colour-coded galaxies by their visual morphology.
We select late-type spirals (Sc+) as being the best tracer of the star-forming main sequence (SFMS), in line with previous works such as \cite{Medling+18}.
We therefore fit the SFMS as a simple linear relationship to the Sc population, using the same procedure as detailed previously in Section \ref{subsec:line_fitting}.

From the SFR-M$_*$ plane, we define quenched galaxies as having a SFR more than 1\,dex below the SFMS.
Despite the arbitrary nature of this threshold, we note that the results in this section are robust against a wide range of selection criteria, even using an offset of just 0.5\,dex.
We apply the 1\,dex selection criteria to our previous \lreeo-based samples, and note that this substantially reduces the sample size available for all our groups, although the effect is not uniform.
Region II now consists of 201 galaxies, whilst Region IV contains just 65 galaxies (previously 316 and 430 galaxies respectively).
Although this limits our analysis somewhat, it is not unexpected.
Galaxies known as ``anaemic spirals", with little ongoing star formation but a visual spiral structure, are known to be only a small fraction of late-type spiral galaxies \citep{van_den_Bergh+76}.

In light of this, we revisit the comparison of slow rotators and Region I galaxies established in Section \ref{subsec:SRs}.
In Fig.\ref{fig:slow_reg_I_hist_comp}, we compare the structural parameters (log$_{10}\left(M_*/M_{\odot}\right)$, $R_e$), and the absolute and residual $\alpha$-enhancement for the quenched sample.
Although there are small shifts in the distributions, and the sample sizes, the overall conclusions are unchanged.
Slow rotators are larger and more massive than Region I galaxies, and correspondingly more $\alpha$-enhanced.
As before though, when comparing the residuals \dafe\ to account for the difference in the mass distributions, we find no such distinction.
In fact, quenched Region I galaxies are enhanced by $\sim0.01$\,dex in \afe\ compared to SRs, although we do not consider this significant due to the uncertainties on the measurements.
These results appear to be consistent with the unquenched sample, although the sample size is too small to draw any firm conclusions.

Considering all classifications, we again analyse the residuals from a global \afe-$\sigma$ relationship, shown in Fig.~\ref{fig:quenched_residuals}.
In the top panel, we refit a linear relationship to the residuals, with the SRs excluded from the fit, following the discussion in Section \ref{subsec:SRs}.
In comparison to Fig.~\ref{fig:sigma_lambda_r_residuals_base}c, the gradient is $\sim$50\% steeper.
Due to the reduction in the sample size, the uncertainty on the fit has also increased, and hence the statistical significance of the relationship is exactly as before.
We note that reusing the fit from Fig.~\ref{fig:sigma_lambda_r_residuals_base}a, compared to refitting the \afe-$\sigma$ relationship without SRs, has no measurable impact when calculating the residuals, and so this can be excluded as a potential source of bias.

In Fig.~\ref{fig:quenched_residuals}b, we consider only the quenched galaxies.
Comparing to the correlation using the full sample, it is immediately clear that the statistical significance of the \dafe-\lreeo\ relation has been substantially reduced.

There are two components to this.
Firstly, the sample size has decreased by over a factor of two, from 1492 galaxies to only 657.
We therefore expect the uncertainty on our fits to increase, and we note that the ratio of the uncertainties scales with the square root of the sample sizes.
Secondly, the removed ``unquenched galaxies'' are likely those with ongoing or recent episodes of star formation.
As such, we anticipate these galaxies to be substantially less $\alpha$-enhanced than those in the ``quenched'' sample, which would also translate to lower \dafe.
By comparing the distribution of galaxies between Fig.~\ref{fig:quenched_residuals}b and Fig.~\ref{fig:quenched_residuals}c, we see that a large proportion of galaxies have been removed from the high \lreeo regime.
The strong gradient for the ``unquenched'' galaxies is therefore the clear cause of the flattening of this \dafe-\lreeo\ relationship in the ``quenched'' sample.

\section{Discussion} \label{sec:discussion}

\subsection{Comparisons to previous studies}

The multitude of methods of calculating the values of age, metallicity and $\alpha$-abundance complicate any comparisons of absolute values across different studies.
Small changes in the method, such as a different set of stellar population models, can cascade into larger systematic offsets in the final results.
Hence, we only consider here the relative trends and scaling relations.

\begin{figure}
    \centering
    \includegraphics[width=\columnwidth]{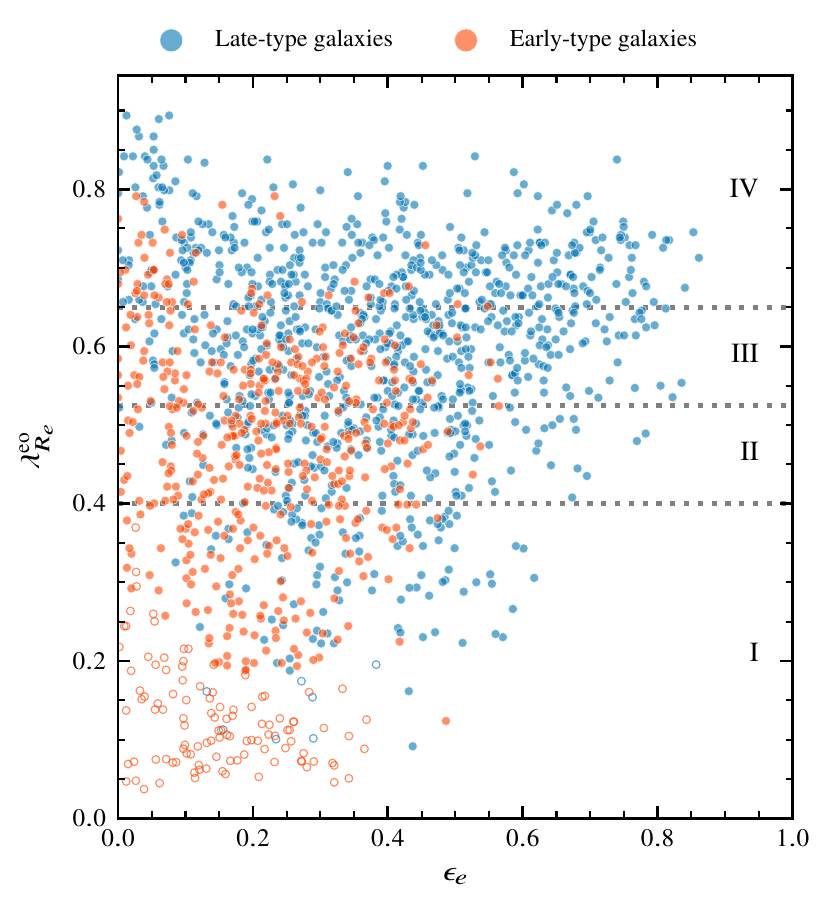}
    \caption{
    The distribution of SGS galaxies in the \lreeo-$\epsilon$ plane, where we have highlighted visually classified ETGs in orange.
    Slow rotators are represented as open circles, whilst all other galaxies are shown as filled markers.
    The definitions of Regions I-IV are reproduced from Table \ref{tab:total_number_of_galaxies}, and are overlaid as dashed lines.
    }
    \label{fig:lambda_ellip_ETG_highlight}
\end{figure}

There are few studies directly linking the kinematic morphologies to stellar populations.
\citetalias{McDermid+15} briefly considered the difference between regular and non-regular rotators (analogous to our Regions I-IV and slow rotators), and found no significant change in the relative abundance of $\alpha$-elements.
This may be partially caused by the limited distribution of ETGs in the \lreeo-$\epsilon$ plane, displayed in Fig.~\ref{fig:lambda_ellip_ETG_highlight} for the SGS.
By focussing on ETGs, \citetalias{McDermid+15} would have also been limited to a similar sample as in Fig.~\ref{fig:quenched_residuals}b.
This is only a weak relationship, and as such, it is highly likely that any correlation between \dafe\ and \lre\ would be drowned out by the scatter in the \afe-$\sigma$ relation, depending on the sample selection.

Perhaps more relevant is \citetalias{Bernardi+19}, which primarily focussed on the radial gradients of galaxies in the MaNGA survey.
The selection used by \citetalias{Bernardi+19} contained almost exclusively ETGs, with the slow rotators separated by means of a simple cut at $\lre=0.2$, rather than the selection criteria first defined by \cite{Cappellari+16} (see Section \ref{sec:kinematic_classification} for more details).
In each stellar mass and size bin, \citetalias{Bernardi+19} found offsets of approximately 0.01-0.05\,dex in \afe\ between fast and slow rotators, depending on the radial aperture size.
This is in broad agreement with our own results, although we note that a similar limitation applies to this as to \citetalias{McDermid+15}, in that results drawn from a binary kinematic classification are strongly dependent on the distribution of galaxies in the \lre-$\epsilon$ plane.

To the best of our knowledge, no other work to date has sought to disentangle the relationship between the spin parameter proxy \lre\ and the abundance of $\alpha$ elements, whilst simultaneously mitigating the effects of ongoing star formation, and so we cannot make any further comparisons here.

\subsection{Physical implications}

In agreement with previous studies, we reproduce the well known relationship between \afe\ and $\sigma$, which indicates that star formation in massive galaxies was quenched relatively faster than in less massive galaxies.

Analysing residuals from the \afe-$\sigma$ relation for the entire sample, we find a statistically significant anti-correlation with the inclination-corrected spin proxy, \lreeo.
This indicates that the kinematic structure of a galaxy plays a role in the quenching mechanism, with disk dominated systems having a longer duration of star formation.
This is in good agreement with \cite{van_de_Sande+18}, who found an analogous relationship between the luminosity-weighted stellar age, and $(V/\sigma)$, the ratio of ordered rotation to random motion, although the different quantities measured mean that our results are not directly comparable.

Considering the background in Section \ref{sec:background}, we suggest that this behaviour is largely consistent with a picture in which classical bulges (CBs) form over considerably shorter periods of time than extended galactic disks.
The decrease in the temporal extent of star formation would lead to a relatively higher \afe\ in the CB compared to the disk.
Considering galaxies at some fixed velocity dispersion, any increase in the bulge-to-disk ratio would correspond to a decrease in the overall rotational support, and hence the measured \lreeo.
The relatively larger bulge would produce a greater fraction of the integrated light, and so a global measurement of \afe\ would be seen to increase.
When taken over a range of velocity dispersion, such a formation scenario would therefore be able to qualitatively explain the observed anti-correlation between \lreeo\ and the residuals \dafe\ shown in Fig.~\ref{fig:sigma_lambda_r_residuals_base}.

However, we note that galaxies containing CBs are not likely to account for the majority of our sample.
Taking a survey of the local 11\,Mpc volume, \cite{Fisher+11} found only 17\%$\pm$10\% of galaxies had an observed CB, relative to 45\%$\pm$12\% with a pseudo-bulge (PB).
This is supported by observational evidence from previous studies in the SGS, with \cite{Barsanti+21} finding 23\% of galaxies contained bulges that were older than their disks (probable CBs), and 34\% where the bulges were younger (probable PBs).
Pseudo-bulges are thought to form out of gas brought to the central regions from the galactic disk, via secular evolution over long timescales \citep{Athanassoula+05}.
In contrast to CBs, which exhibit a wide range of star-formation rates, PBs are almost universally star-forming \citep{Luo+20}.
Integrated over a galaxy, this longer extent of star formation in PBs would suggest a noticeably lower measurement of \afe\ compared to CBs.
Since PBs do not have substantially different star-formation histories to their surrounding disks, we do not expect the host galaxies to show any clear relationship between their degree of rotational support, and their $\alpha$-enhancement.

From Fig.~\ref{fig:sfms_linefit_sb_col_vis_morph}, we can see that our sample contains a considerable fraction of actively star-forming galaxies.
Whilst \cite{Luo+20} demonstrated that CBs themselves have a wide-range of star-formation rates, a large share of their host galaxies are still star-forming, depending on the morphology and environment \citep{Mishra+19}.
As such, when investigating the quenched sample, we remove a considerable proportion of galaxies for which we expect \lreeo\ and \dafe\ to be correlated, leading to the relationship displayed in Fig.~\ref{fig:quenched_residuals}b.
By comparison, our unquenched sample most probably consists of a mixture of CBs and PBs.
Whilst Fig.~\ref{fig:quenched_residuals}c demonstrates a clear anti-correlation between the degree of rotational support and $\alpha$-enhancement, we suggest that the considerable scatter is largely driven by galaxies hosting PBs.

There is considerable support in existing literature for differing quenching mechanisms as a function of kinematic morphology, such as \cite{Smethurst+15}.
In particular, for disk-dominated galaxies, which would have high \lreeo, \cite{Smethurst+15} found that the dominant quenching mechanisms were those taking place over long timescales (defined as $\tau>2$\,Gyr).
These are likely to be secular processes, i.e. those internal to the galaxy, such as bar formation in spiral galaxies funnelling gas towards central PBs \citep{Cheung+13}.
By comparison, the dominant mechanisms for smooth galaxies (those with lower \lreeo) are those taking place over the shortest time-scales ($\tau<1$\,Gyr), such as dry major mergers, perhaps in combination with AGN feedback \citep{Springel+05}.
We would expect the relatively longer quenching timescales for disk-dominated systems to lead to a lower $\alpha$-abundance.
However, whilst there is some evidence for this in Section \ref{sec:quenched_galaxies}, this is not statistically significant enough for us to draw any firm conclusions.
In particular, we cannot conclude whether the observed relationship in Fig.~\ref{fig:quenched_residuals}b is due to morphological-dependent quenching mechanisms, or a weak residual effect from galaxies containing CBs.
We suggest that further study with a larger sample size would be very helpful, especially in order to clarify the statistical significance of this relation.
Disentangling the relative contribution of bulges to the overall $\alpha$-enhancement of galaxies requires a spatially-resolved view, and we intend to explore this effect in more detail in a future paper.

Although there is only weak evidence for this \dafe-\lreeo\ relationship persisting in the quenched sample, we show in Fig.~\ref{fig:quenched_residuals}c that the relationship in the full sample is largely driven by unquenched galaxies.
To estimate the physical implications of this result, we can use the empirical formulae of \cite{de_la_Rosa+11}.
These were obtained by utilising full-spectrum fitting in a sample of elliptical galaxies to determine non-parametric star formation histories, and comparing to the abundance ratios derived from line indices, using the same models as we have in this paper \citep{Thomas+10}.
One of these formulae considers the half-mass formation time ($T_{\rm{M/2}}$) as a function of $\alpha$-abundance,
$$
    T_{\rm{M/2}} (\rm{Gyr})=-15.3\afe+5.2.
$$
Looking at the relationship in Fig.~\ref{fig:quenched_residuals}c, we can therefore predict how $T_{\rm{M/2}}$ varies at some fixed velocity dispersion,
$$
    \left.\Delta T_{\rm{M/2}} (\rm{Gyr})\right\vert_{\rm{unquenched}}=(-1.85\pm0.23)\Delta\lreeo.
$$
As an example, considering the median galaxy in each group, we would expect a Region IV galaxy to have a half-mass formation time $\sim$0.7\,Gyr longer than a galaxy in Region I.
To put this in context, at a fixed velocity dispersion of 200\,km\,s$^{-1}$, considering the global relationship in Fig.~\ref{fig:sigma_lambda_r_residuals_base}a, a Region I galaxy would have $T_{\rm{M/2}}=0.51$\,Gyr, whereas a Region IV galaxy would have $T_{\rm{M/2}}=1.26$\,Gyr.

\begin{figure}
    \centering
    \includegraphics[width=\columnwidth]{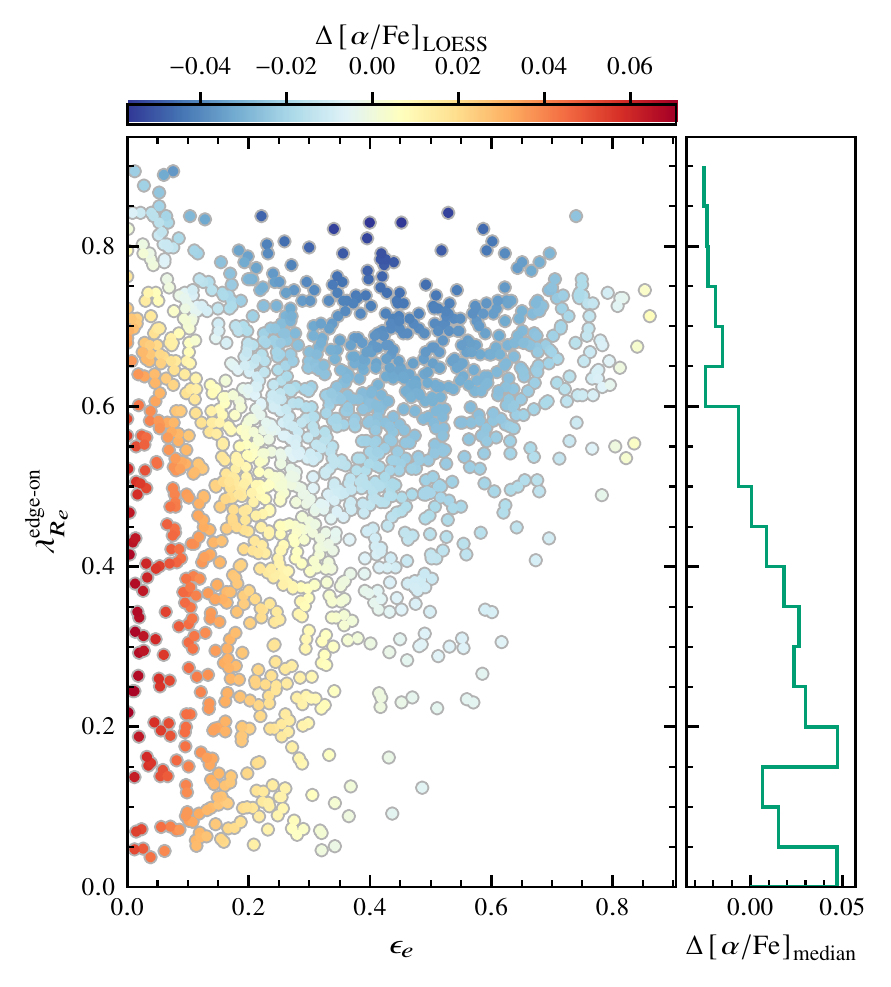}
    \caption{
    The distribution of SGS galaxies in the \lreeo-$\epsilon$ plane.
    Galaxies are coloured according to their residual $\alpha$-enhancement, measured relative to the best-fit relation displayed in Fig.~\ref{fig:sigma_lambda_r_residuals_base}.
    These values have been smoothed using the locally weighted regression technique (LOESS) of \protect\cite{Cappellari+13b}.
    The stepped function on the right represents the residuals prior to smoothing, with the median \dafe\ taken from bins of 0.05 \lreeo.
    }
    \label{fig:residual_LOESS}
\end{figure}

Turning our attention to the slow rotators, the offsets in the distributions in Fig.~\ref{fig:slow_reg_I_hist_comp} support the concept that slow and fast rotators form distinct populations, as far as their structural parameters are concerned.
However, we find little evidence for such a binary classification in the stellar populations, with the slow rotators (SRs) having a residual $\alpha$-enhancement consistent with Region I galaxies.
Considering similar population parameters such as the mean stellar age, we note that studies such as \cite{van_de_Sande+18} have previously shown a smooth variation of stellar populations across the slow/fast rotator boundary.

Therefore, the unexpected result here is that the residual $\alpha$-enhancement appears to level off for $\lreeo<0.4$.
We illustrate this more clearly in Fig.~\ref{fig:residual_LOESS}, where we show the distribution of residuals in the \lreeo-$\epsilon$ plane, after local smoothing using the LOESS algorithm \citep{Cappellari+13b}.
Whilst there is a clear trend towards lower values of \dafe\ for $\lreeo>0.4$, there is no appreciable difference in the residuals below this point.
This may be partially due to the difficulty in determining a clean separation between the two groups.
\cite{van_de_Sande+21a} expands upon this problem of how to accurately determine distinct kinematic classifications in considerable detail, although it is clear that any solution contains a certain degree of compromise. 
Simplistically, we can see in Fig.~\ref{fig:SR_selection} the density of galaxies in \lre-$\epsilon$ space around the cutoff.
However, from Fig.~\ref{fig:sigma_lambda_r_residuals_base}c, we can also see that the distribution of SRs contains no galaxies substantially higher in \dafe\ than those in Region I.
This cannot be changed by altering $\lambda_{R_{\rm{start}}}$ or the gradient of the dividing line, and so we reject this selection effect as the driver behind our results.

An additional factor that must be considered is that the SAMI sample is known to contain a non-zero number of galaxies with counter-rotating disks \citep{Rawlings+20}, also referred to as 2$\sigma$ galaxies.
When integrated over a wide enough aperture, these will have a low measured spin proxy due to the summation of the separate disk components, and may be misidentified as SRs.
Although most of these galaxies have $\epsilon>0.4$, and would be caught by the selection criteria introduced in \cite{Cappellari+16}, we cannot rule out that this may influence the exact relationship.

\begin{figure}
    \centering
    \includegraphics[width=\columnwidth]{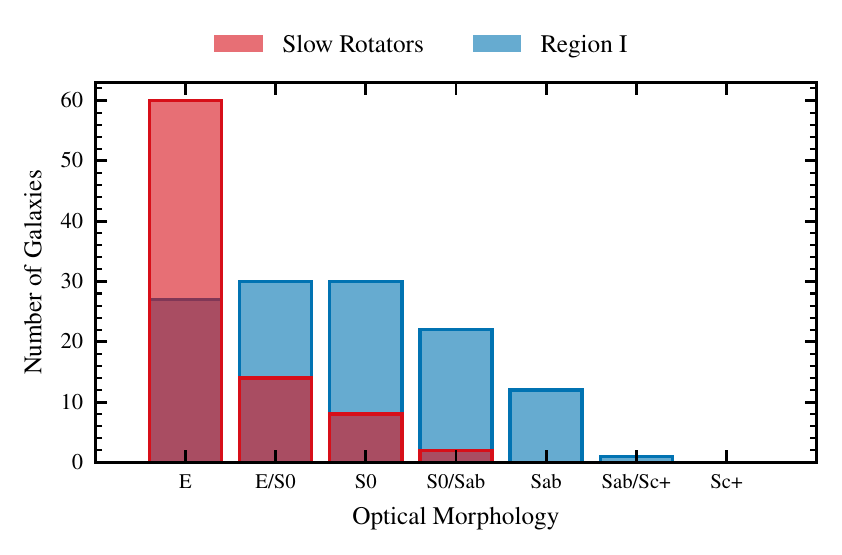}
    \caption{
    The morphological distribution of slow rotators and Region I galaxies in our quenched sample.
    Slow rotators are almost exclusively ellipticals, whereas Region I galaxies show a much more balanced distribution of optical morphologies.
    }
    \label{fig:SR_Reg_I_morph}
\end{figure}

As before for the general anti-correlation described in Fig.~\ref{fig:sigma_lambda_r_residuals_base}, the similarity between the residual $\alpha$-enhancement distributions of SRs and Region I galaxies cannot be easily attributed to differing stellar populations across bulge and disk components.
Whilst there are bulge-disk decompositions for the SGS, these have been limited so far to subsamples, such as the lenticular cluster sample explored in \cite{Barsanti+21} and the GAMA regions in Lah \textit{et al.} (in prep.), or with stringent quality cuts \citep{Oh+20}.
We instead look at the optical morphological classifications of \cite{Cortese+16}, presented in Fig.~\ref{fig:SR_Reg_I_morph} for our quenched sample.
Whilst SRs are heavily bulge-dominated systems, with $\sim$two-thirds unambiguously classified as ellipticals, Region I galaxies show a much greater variation in their morphological classifications.
This is in line with the other structural parameters explored in Fig.~\ref{fig:slow_reg_I_hist_comp}.
Hence, despite the limitations discussed previously, we maintain that the similar residual $\alpha$-enhancement of SRs and Region I galaxies in our sample is a real phenomenon, and one that we hope can be replicated across other spectroscopic surveys.

To explain the likely reasons behind this phenomenon, we explore the slow rotator formation scenarios espoused by \cite{Naab+14}.
Using a slightly different selection criteria, we see that there are two likely pathways that agree with observations:
\begin{itemize}
    \item Class C: Late ($z\lesssim1$), gas-rich major mergers.
    The high fraction of stars formed in the galaxy, relative to those accreted through a merger, would likely result in a low integrated \afe\ ratio.
    \item Class F: Repeated minor mergers, from at least $z\sim2$.
    These would likely have significantly higher \afe\ than Class C galaxies.
\end{itemize}
We can see already that the two formation scenarios would likely result in a broad spread of \afe\ relative to fast rotators, and consider the possible mix of Class C and F galaxies as the most probable cause for slow rotators not following the residual anti-correlation between \dafe\ and \lreeo.
We further speculate that under the formation pathway of Class F galaxies, a recent minor merger with a late-type galaxy may also depress the integrated \afe\ ratio, and result in measurements closer to solar values.
In addition, the sample used by \cite{Naab+14} predicts both an increase in the median stellar mass and radius of slow rotators relative to fast rotators, in good agreement with our results from Fig.~\ref{fig:slow_reg_I_hist_comp}.

More recently, these results have been reinforced by \cite{Lagos+21}, who investigated slow rotator formation scenarios in the EAGLE simulation.
They explicitly computed the expected chemical abundances for each assembly history, as a function of galaxy radius.
All slow rotators that formed via mergers were $\alpha$-enhanced, relative to main sequence galaxies, whereas those that formed without mergers, and hence had little disruption to ongoing star formation, were predicted to have considerably lower \afe\ towards the centre.
Similarly, investigating slow rotators that have had any merger in the last 10~Gyrs, those with gas-rich mergers were $\sim0.1$~dex lower in \afe\ than the average, in agreement with the expectations from \cite{Naab+14}.
Both the ``no merger'' and ``gas-rich merger'' scenarios could explain the comparable residual $\alpha$-enhancement of SRs to Region I galaxies.
Further work on this area is clearly needed however, as the spatial information in the kinematic maps may be able to reveal which formation pathway is dominant in this sample. 

\section{Conclusions} \label{sec:conclusions}

We utilise measurements of \afe\ from \cite{Watson+21} and \lreeo from \cite{van_de_Sande+21a} to analyse galaxies in the SAMI Galaxy Survey. Based on our analysis of the kinematic properties across the sample, we make the following observations:
\begin{enumerate}
    \item For all 1492 galaxies with measurements of \lre, we fit a global \afe-$\sigma$ relation to obtain the result
    \begin{equation*}
         \afe=(0.395\pm0.010)\rm{log}_{10}\left(\sigma\right)-(0.627\pm0.002).
    \end{equation*}
    We find that the residuals \dafe\ display a strong anti-correlation with the inclination corrected \lreeo,
    \begin{equation*}
        \dafe=(-0.057\pm0.008)\lreeo+(0.020\pm0.003).
    \end{equation*}
    If we use \lre\ rather than \lreeo, the gradient of the correlation is over 50\% higher.
    \item We isolate the quenched population based on a 1\,dex offset in star formation rate from the star-forming main sequence.
    This substantially reduces the statistical significance of this anti-correlation, which now has a gradient $m=-0.038\pm0.020$.
    However, the persistence of this relation, whilst weak, demonstrates that the offset cannot unambiguously be attributed to ongoing star formation.
    \item Unquenched galaxies demonstrate a much more significant relationship, 
    \begin{equation*}
        \dafe=(-0.121\pm0.015)\lreeo+(0.046\pm0.009).
    \end{equation*}
    \item This anti-correlation would therefore imply a link between the duration of star formation and the angular momentum of the galaxy. 
    Utilising the empirical formulae of \cite{de_la_Rosa+11}, we estimate the half-mass formation time of unquenched galaxies in Region IV to be extended by $\sim$0.7\,Gyr relative to those in Region I.
    \begin{equation}
        \Delta T_{\rm{M/2}} (\rm{Gyr})=(-1.85\pm0.23)\Delta\lreeo
    \end{equation}
    We hope to further explore this link in future studies.
    \item Comparing slow rotators to the lowest \lreeo fast rotators (Region I), we find that the difference in $\alpha$-enhancement is determined to first order by $\sigma$.
    Accounting for this, by looking at the residuals \dafe, we find the slow rotators and Region I galaxies are indistinguishable.
    The star formation histories of these two groups ($\lreeo\lessapprox0.4$) are therefore similar, despite a substantial difference in their structure and kinematics.
    This contrasts with galaxies in Regions II-IV, which show increasingly extended star formation histories, correlated with their degree of rotational support.
    
\end{enumerate}

\section*{Acknowledgements}

The  SAMI Galaxy Survey is based on observations made at the Anglo-Australian Telescope, and was developed jointly by the University of Sydney and the Australian Astronomical Observatory.
The SAMI input catalogue is based on data taken from the Sloan Digital Sky Survey, the GAMA Survey, and the VST ATLAS Survey.
The SAMI Galaxy Survey is supported by the Australian Research Council Centre of Excellence for All Sky Astrophysics in 3 Dimensions (ASTRO 3D), through project number CE170100013, the Australian Research Council Centre of Excellence for All-sky Astrophysics (CAASTRO), through project number CE110001020, and other participating institutions. 
The SAMI Galaxy Survey website is \href{http://sami-survey.org/}{http://sami-survey.org/}.

PJW and RLD acknowledge travel and computer grants from Christ Church, Oxford and support from the Oxford Hintze Centre for Astrophysical Surveys which is funded by the Hintze Family Charitable Foundation.
RLD is also supported by the Science and Technology Facilities Council grant numbers ST/H002456/1, ST/K00106X/1 and ST/J002216/1.
JvdS acknowledges support of an Australian Research Council Discovery Early Career Research Award (DE200100461) funded by the Australian Government.
SB acknowledges funding support from the Australian Research Council through a Future Fellowship (FT140101166).
FDE acknowledges funding through the ERC Advanced grant 695671 ``QUENCH'', the H2020 ERC Consolidator Grant 683184 and support by the Science and Technology Facilities Council (STFC). 
BG is the recipient of an Australian Research Council Future Fellowship (FT140101202).
NS acknowledges support of an Australian Research Council Discovery Early Career Research Award (DE190100375) funded by the Australian Government and a University of Sydney Postdoctoral Research Fellowship.
JBH is supported by an Australian Research Council Laureate Fellowship (FL140100278). 
The SAMI instrument was funded by Bland-Hawthorn’s former Federation Fellowship (FF0776384), an Australian Research Council Linkage Infrastructure, Equipment and Facilities grant (LE130100198; PI Bland-Hawthorn) and funding from the Anglo-Australian Observatory.
JJB acknowledges support of an Australian Research Council Future Fellowship (FT180100231).
MSO acknowledges funding support from the Australian Research Council through a Future Fellowship (FT140100255).

%%%%%%%%%%%%%%%%%%%%%%%%%%%%%%%%%%%%%%%%%%%%%%%%%%
\section*{Data Availability}

All data used in this work are publicly available from Astronomical Optics' Data Central service at \href{https://datacentral.org.au/}{https://datacentral.org.au/} as part of the SAMI Data Release 3 \citep{Croom+21}.

%%%%%%%%%%%%%%%%%%%% REFERENCES %%%%%%%%%%%%%%%%%%

% The best way to enter references is to use BibTeX:

\bibliographystyle{mnras}
\bibliography{full_list.bib}

\begin{thebibliography}{}
\makeatletter
\relax
\def\mn@urlcharsother{\let\do\@makeother \do\$\do\&\do\#\do\^\do\_\do\%\do\~}
\def\mn@doi{\begingroup\mn@urlcharsother \@ifnextchar [ {\mn@doi@}
  {\mn@doi@[]}}
\def\mn@doi@[#1]#2{\def\@tempa{#1}\ifx\@tempa\@empty \href
  {http://dx.doi.org/#2} {doi:#2}\else \href {http://dx.doi.org/#2} {#1}\fi
  \endgroup}
\def\mn@eprint#1#2{\mn@eprint@#1:#2::\@nil}
\def\mn@eprint@arXiv#1{\href {http://arxiv.org/abs/#1} {{\tt arXiv:#1}}}
\def\mn@eprint@dblp#1{\href {http://dblp.uni-trier.de/rec/bibtex/#1.xml}
  {dblp:#1}}
\def\mn@eprint@#1:#2:#3:#4\@nil{\def\@tempa {#1}\def\@tempb {#2}\def\@tempc
  {#3}\ifx \@tempc \@empty \let \@tempc \@tempb \let \@tempb \@tempa \fi \ifx
  \@tempb \@empty \def\@tempb {arXiv}\fi \@ifundefined
  {mn@eprint@\@tempb}{\@tempb:\@tempc}{\expandafter \expandafter \csname
  mn@eprint@\@tempb\endcsname \expandafter{\@tempc}}}

\bibitem[\protect\citeauthoryear{{Ahn} et~al.,}{{Ahn} et~al.}{2012}]{Ahn+12}
{Ahn} C.~P.,  et~al., 2012, \mn@doi [\apjs] {10.1088/0067-0049/203/2/21}, \href
  {http://adsabs.harvard.edu/abs/2012ApJS..203...21A} {203, 21}

\bibitem[\protect\citeauthoryear{{Allen} et~al.,}{{Allen}
  et~al.}{2015}]{Allen+15}
{Allen} J.~T.,  et~al., 2015, \mn@doi [\mnras] {10.1093/mnras/stu2057}, \href
  {https://ui.adsabs.harvard.edu/abs/2015MNRAS.446.1567A} {446, 1567}

\bibitem[\protect\citeauthoryear{{Annibali}, {Gr{\"u}tzbauch}, {Rampazzo},
  {Bressan}  \& {Zeilinger}}{{Annibali} et~al.}{2011}]{Annibali+11}
{Annibali} F.,  {Gr{\"u}tzbauch} R.,  {Rampazzo} R.,  {Bressan} A.,
  {Zeilinger} W.~W.,  2011, \mn@doi [\aap] {10.1051/0004-6361/201015635}, \href
  {https://ui.adsabs.harvard.edu/abs/2011A&A...528A..19A} {528, A19}

\bibitem[\protect\citeauthoryear{{Athanassoula}}{{Athanassoula}}{2005}]{Athanassoula+05}
{Athanassoula} E.,  2005, \mn@doi [\mnras] {10.1111/j.1365-2966.2005.08872.x},
  \href {https://ui.adsabs.harvard.edu/abs/2005MNRAS.358.1477A} {358, 1477}

\bibitem[\protect\citeauthoryear{{Barsanti} et~al.,}{{Barsanti}
  et~al.}{2021}]{Barsanti+21}
{Barsanti} S.,  et~al., 2021, \mn@doi [\apj] {10.3847/1538-4357/abc956}, \href
  {https://ui.adsabs.harvard.edu/abs/2021ApJ...906..100B} {906, 100}

\bibitem[\protect\citeauthoryear{{Bender}, {Burstein}  \& {Faber}}{{Bender}
  et~al.}{1992}]{Bender+92}
{Bender} R.,  {Burstein} D.,   {Faber} S.~M.,  1992, \mn@doi [\apj]
  {10.1086/171940}, \href
  {https://ui.adsabs.harvard.edu/abs/1992ApJ...399..462B} {399, 462}

\bibitem[\protect\citeauthoryear{{Bernardi}, {Dom{\'\i}nguez S{\'a}nchez},
  {Brownstein}, {Drory}  \& {Sheth}}{{Bernardi} et~al.}{2019}]{Bernardi+19}
{Bernardi} M.,  {Dom{\'\i}nguez S{\'a}nchez} H.,  {Brownstein} J.~R.,  {Drory}
  N.,   {Sheth} R.~K.,  2019, \mn@doi [\mnras] {10.1093/mnras/stz2413}, \href
  {https://ui.adsabs.harvard.edu/abs/2019MNRAS.489.5633B} {489, 5633}

\bibitem[\protect\citeauthoryear{{Bland-Hawthorn} et~al.,}{{Bland-Hawthorn}
  et~al.}{2011}]{Bland_Hawthorn+11}
{Bland-Hawthorn} J.,  et~al., 2011, \mn@doi [Optics Express]
  {10.1364/OE.19.002649}, \href
  {http://adsabs.harvard.edu/abs/2011OExpr..19.2649B} {19, 2649}

\bibitem[\protect\citeauthoryear{{Bryant}, {Bland-Hawthorn}, {Fogarty},
  {Lawrence}  \& {Croom}}{{Bryant} et~al.}{2014}]{Bryant+14}
{Bryant} J.~J.,  {Bland-Hawthorn} J.,  {Fogarty} L.~M.~R.,  {Lawrence} J.~S.,
  {Croom} S.~M.,  2014, \mn@doi [\mnras] {10.1093/mnras/stt2254}, \href
  {http://adsabs.harvard.edu/abs/2014MNRAS.438..869B} {438, 869}

\bibitem[\protect\citeauthoryear{{Bryant} et~al.,}{{Bryant}
  et~al.}{2015}]{Bryant+15}
{Bryant} J.~J.,  et~al., 2015, \mn@doi [\mnras] {10.1093/mnras/stu2635}, \href
  {https://ui.adsabs.harvard.edu/abs/2015MNRAS.447.2857B} {447, 2857}

\bibitem[\protect\citeauthoryear{{Bundy} et~al.,}{{Bundy}
  et~al.}{2015}]{Bundy+15}
{Bundy} K.,  et~al., 2015, \mn@doi [\apj] {10.1088/0004-637X/798/1/7}, \href
  {https://ui.adsabs.harvard.edu/abs/2015ApJ...798....7B} {798, 7}

\bibitem[\protect\citeauthoryear{{Cappellari}}{{Cappellari}}{2002}]{Cappellari+02}
{Cappellari} M.,  2002, \mn@doi [\mnras] {10.1046/j.1365-8711.2002.05412.x},
  \href {https://ui.adsabs.harvard.edu/abs/2002MNRAS.333..400C} {333, 400}

\bibitem[\protect\citeauthoryear{{Cappellari}}{{Cappellari}}{2016}]{Cappellari+16}
{Cappellari} M.,  2016, \mn@doi [\araa] {10.1146/annurev-astro-082214-122432},
  \href {https://ui.adsabs.harvard.edu/abs/2016ARA&A..54..597C} {54, 597}

\bibitem[\protect\citeauthoryear{{Cappellari}}{{Cappellari}}{2017}]{Cappellari+17}
{Cappellari} M.,  2017, \mn@doi [\mnras] {10.1093/mnras/stw3020}, \href
  {https://ui.adsabs.harvard.edu/abs/2017MNRAS.466..798C} {466, 798}

\bibitem[\protect\citeauthoryear{{Cappellari} et~al.,}{{Cappellari}
  et~al.}{2011}]{Cappellari+11}
{Cappellari} M.,  et~al., 2011, \mn@doi [\mnras]
  {10.1111/j.1365-2966.2010.18174.x}, \href
  {https://ui.adsabs.harvard.edu/abs/2011MNRAS.413..813C} {413, 813}

\bibitem[\protect\citeauthoryear{{Cappellari} et~al.,}{{Cappellari}
  et~al.}{2013}]{Cappellari+13b}
{Cappellari} M.,  et~al., 2013, \mn@doi [\mnras] {10.1093/mnras/stt644}, \href
  {https://ui.adsabs.harvard.edu/abs/2013MNRAS.432.1862C} {432, 1862}

\bibitem[\protect\citeauthoryear{{Cenarro}, {Cardiel}, {Gorgas}, {Peletier},
  {Vazdekis}  \& {Prada}}{{Cenarro} et~al.}{2001}]{Cenarro+01}
{Cenarro} A.~J.,  {Cardiel} N.,  {Gorgas} J.,  {Peletier} R.~F.,  {Vazdekis}
  A.,   {Prada} F.,  2001, \mn@doi [\mnras] {10.1046/j.1365-8711.2001.04688.x},
  \href {https://ui.adsabs.harvard.edu/abs/2001MNRAS.326..959C} {326, 959}

\bibitem[\protect\citeauthoryear{{Chabrier}}{{Chabrier}}{2003}]{Chabrier+03}
{Chabrier} G.,  2003, \mn@doi [\pasp] {10.1086/376392}, \href
  {https://ui.adsabs.harvard.edu/abs/2003PASP..115..763C} {115, 763}

\bibitem[\protect\citeauthoryear{{Cheung} et~al.,}{{Cheung}
  et~al.}{2013}]{Cheung+13}
{Cheung} E.,  et~al., 2013, \mn@doi [\apj] {10.1088/0004-637X/779/2/162}, \href
  {https://ui.adsabs.harvard.edu/abs/2013ApJ...779..162C} {779, 162}

\bibitem[\protect\citeauthoryear{{Cortese} et~al.,}{{Cortese}
  et~al.}{2016}]{Cortese+16}
{Cortese} L.,  et~al., 2016, \mn@doi [\mnras] {10.1093/mnras/stw1891}, \href
  {https://ui.adsabs.harvard.edu/abs/2016MNRAS.463..170C} {463, 170}

\bibitem[\protect\citeauthoryear{{Croom} et~al.,}{{Croom}
  et~al.}{2012}]{Croom+12}
{Croom} S.~M.,  et~al., 2012, \mn@doi [\mnras]
  {10.1111/j.1365-2966.2011.20365.x}, \href
  {https://ui.adsabs.harvard.edu/abs/2012MNRAS.421..872C} {421, 872}

\bibitem[\protect\citeauthoryear{{Croom} et~al.,}{{Croom}
  et~al.}{2021}]{Croom+21}
{Croom} S.~M.,  et~al., 2021, \mn@doi [\mnras] {10.1093/mnras/stab229}, \href
  {https://ui.adsabs.harvard.edu/abs/2021MNRAS.tmp..291C} {}

\bibitem[\protect\citeauthoryear{{D'Eugenio} et~al.,}{{D'Eugenio}
  et~al.}{2021}]{D'Eugenio+21}
{D'Eugenio} F.,  et~al., 2021, \mn@doi [\mnras] {10.1093/mnras/stab1146}, \href
  {https://ui.adsabs.harvard.edu/abs/2021MNRAS.tmp.1133D} {}

\bibitem[\protect\citeauthoryear{{Davies}, {Sadler}  \& {Peletier}}{{Davies}
  et~al.}{1993}]{Davies+93}
{Davies} R.~L.,  {Sadler} E.~M.,   {Peletier} R.~F.,  1993, \mn@doi [\mnras]
  {10.1093/mnras/262.3.650}, \href
  {https://ui.adsabs.harvard.edu/abs/1993MNRAS.262..650D} {262, 650}

\bibitem[\protect\citeauthoryear{{Driver} et~al.,}{{Driver}
  et~al.}{2011}]{Driver+11}
{Driver} S.~P.,  et~al., 2011, \mn@doi [\mnras]
  {10.1111/j.1365-2966.2010.18188.x}, \href
  {https://ui.adsabs.harvard.edu/abs/2011MNRAS.413..971D} {413, 971}

\bibitem[\protect\citeauthoryear{{Eggen}, {Lynden-Bell}  \& {Sandage}}{{Eggen}
  et~al.}{1962}]{Eggen+62}
{Eggen} O.~J.,  {Lynden-Bell} D.,   {Sandage} A.~R.,  1962, \mn@doi [\apj]
  {10.1086/147433}, \href
  {https://ui.adsabs.harvard.edu/abs/1962ApJ...136..748E} {136, 748}

\bibitem[\protect\citeauthoryear{{Emsellem}, {Monnet}  \& {Bacon}}{{Emsellem}
  et~al.}{1994}]{Emsellem+94}
{Emsellem} E.,  {Monnet} G.,   {Bacon} R.,  1994, \aap, \href
  {https://ui.adsabs.harvard.edu/abs/1994A&A...285..723E} {285, 723}

\bibitem[\protect\citeauthoryear{{Emsellem} et~al.,}{{Emsellem}
  et~al.}{2007}]{Emsellem+07}
{Emsellem} E.,  et~al., 2007, \mn@doi [\mnras]
  {10.1111/j.1365-2966.2007.11752.x}, \href
  {https://ui.adsabs.harvard.edu/abs/2007MNRAS.379..401E} {379, 401}

\bibitem[\protect\citeauthoryear{{Emsellem} et~al.,}{{Emsellem}
  et~al.}{2011}]{Emsellem+11}
{Emsellem} E.,  et~al., 2011, \mn@doi [\mnras]
  {10.1111/j.1365-2966.2011.18496.x}, \href
  {https://ui.adsabs.harvard.edu/abs/2011MNRAS.414..888E} {414, 888}

\bibitem[\protect\citeauthoryear{{Falc{\'o}n-Barroso},
  {S{\'a}nchez-Bl{\'a}zquez}, {Vazdekis}, {Ricciardelli}, {Cardiel}, {Cenarro},
  {Gorgas}  \& {Peletier}}{{Falc{\'o}n-Barroso} et~al.}{2011}]{Vazdekis+11}
{Falc{\'o}n-Barroso} J.,  {S{\'a}nchez-Bl{\'a}zquez} P.,  {Vazdekis} A.,
  {Ricciardelli} E.,  {Cardiel} N.,  {Cenarro} A.~J.,  {Gorgas} J.,
  {Peletier} R.~F.,  2011, \mn@doi [\aap] {10.1051/0004-6361/201116842}, \href
  {https://ui.adsabs.harvard.edu/abs/2011A&A...532A..95F} {532, A95}

\bibitem[\protect\citeauthoryear{{Fisher} \& {Drory}}{{Fisher} \&
  {Drory}}{2011}]{Fisher+11}
{Fisher} D.~B.,  {Drory} N.,  2011, \mn@doi [\apjl]
  {10.1088/2041-8205/733/2/L47}, \href
  {https://ui.adsabs.harvard.edu/abs/2011ApJ...733L..47F} {733, L47}

\bibitem[\protect\citeauthoryear{{Franx} \& {Illingworth}}{{Franx} \&
  {Illingworth}}{1990}]{Franx+90}
{Franx} M.,  {Illingworth} G.,  1990, \mn@doi [\apjl] {10.1086/185791}, \href
  {https://ui.adsabs.harvard.edu/abs/1990ApJ...359L..41F} {359, L41}

\bibitem[\protect\citeauthoryear{{Graham} et~al.,}{{Graham}
  et~al.}{2018}]{Graham+18}
{Graham} M.~T.,  et~al., 2018, \mn@doi [\mnras] {10.1093/mnras/sty504}, \href
  {https://ui.adsabs.harvard.edu/abs/2018MNRAS.477.4711G} {477, 4711}

\bibitem[\protect\citeauthoryear{{Green} et~al.,}{{Green}
  et~al.}{2018}]{Green+18}
{Green} A.~W.,  et~al., 2018, \mn@doi [\mnras] {10.1093/mnras/stx3135}, \href
  {https://ui.adsabs.harvard.edu/abs/2018MNRAS.475..716G} {475, 716}

\bibitem[\protect\citeauthoryear{{Greggio} \& {Renzini}}{{Greggio} \&
  {Renzini}}{1983}]{Greggio+83}
{Greggio} L.,  {Renzini} A.,  1983, \aap, \href
  {https://ui.adsabs.harvard.edu/abs/1983A&A...118..217G} {118, 217}

\bibitem[\protect\citeauthoryear{{Harborne}, {van de Sande}, {Cortese},
  {Power}, {Robotham}, {Lagos}  \& {Croom}}{{Harborne}
  et~al.}{2020}]{Harborne+20a}
{Harborne} K.~E.,  {van de Sande} J.,  {Cortese} L.,  {Power} C.,  {Robotham}
  A.~S.~G.,  {Lagos} C.~D.~P.,   {Croom} S.,  2020, \mn@doi [\mnras]
  {10.1093/mnras/staa1847}, \href
  {https://ui.adsabs.harvard.edu/abs/2020MNRAS.497.2018H} {497, 2018}

\bibitem[\protect\citeauthoryear{{Kennicutt}, {Tamblyn}  \&
  {Congdon}}{{Kennicutt} et~al.}{1994}]{Kennicutt+94}
{Kennicutt} Robert~C. J.,  {Tamblyn} P.,   {Congdon} C.~E.,  1994, \mn@doi
  [\apj] {10.1086/174790}, \href
  {https://ui.adsabs.harvard.edu/abs/1994ApJ...435...22K} {435, 22}

\bibitem[\protect\citeauthoryear{{Krajnovi{\'c}} et~al.,}{{Krajnovi{\'c}}
  et~al.}{2011}]{Krajnovic+11}
{Krajnovi{\'c}} D.,  et~al., 2011, \mn@doi [\mnras]
  {10.1111/j.1365-2966.2011.18560.x}, \href
  {https://ui.adsabs.harvard.edu/abs/2011MNRAS.414.2923K} {414, 2923}

\bibitem[\protect\citeauthoryear{{Krajnovi{\'c}} et~al.,}{{Krajnovi{\'c}}
  et~al.}{2020}]{Krajnovic+20}
{Krajnovi{\'c}} D.,  et~al., 2020, \mn@doi [\aap]
  {10.1051/0004-6361/201937040}, \href
  {https://ui.adsabs.harvard.edu/abs/2020A&A...635A.129K} {635, A129}

\bibitem[\protect\citeauthoryear{{Lagos}, {Emsellem}, {van de Sande},
  {Harborne}, {Cortese}, {Davison}, {Foster}  \& {Wright}}{{Lagos}
  et~al.}{2021}]{Lagos+21}
{Lagos} C. d.~P.,  {Emsellem} E.,  {van de Sande} J.,  {Harborne} K.~E.,
  {Cortese} L.,  {Davison} T.,  {Foster} C.,   {Wright} R.~J.,  2021, \mn@doi
  [\mnras] {10.1093/mnras/stab3128}, \href
  {https://ui.adsabs.harvard.edu/abs/2021MNRAS.tmp.2880L} {}

\bibitem[\protect\citeauthoryear{{Larson}}{{Larson}}{1974}]{Larson+74}
{Larson} R.~B.,  1974, \mn@doi [\mnras] {10.1093/mnras/166.3.585}, \href
  {https://ui.adsabs.harvard.edu/abs/1974MNRAS.166..585L} {166, 585}

\bibitem[\protect\citeauthoryear{{Luo} et~al.,}{{Luo} et~al.}{2020}]{Luo+20}
{Luo} Y.,  et~al., 2020, \mn@doi [\mnras] {10.1093/mnras/staa328}, \href
  {https://ui.adsabs.harvard.edu/abs/2020MNRAS.493.1686L} {493, 1686}

\bibitem[\protect\citeauthoryear{{McDermid} et~al.,}{{McDermid}
  et~al.}{2015}]{McDermid+15}
{McDermid} R.~M.,  et~al., 2015, \mn@doi [\mnras] {10.1093/mnras/stv105}, \href
  {https://ui.adsabs.harvard.edu/abs/2015MNRAS.448.3484M} {448, 3484}

\bibitem[\protect\citeauthoryear{{Medling} et~al.,}{{Medling}
  et~al.}{2018}]{Medling+18}
{Medling} A.~M.,  et~al., 2018, \mn@doi [\mnras] {10.1093/mnras/sty127}, \href
  {https://ui.adsabs.harvard.edu/abs/2018MNRAS.475.5194M} {475, 5194}

\bibitem[\protect\citeauthoryear{{Mishra}, {Wadadekar}  \& {Barway}}{{Mishra}
  et~al.}{2019}]{Mishra+19}
{Mishra} P.~K.,  {Wadadekar} Y.,   {Barway} S.,  2019, \mn@doi [\mnras]
  {10.1093/mnras/stz1621}, \href
  {https://ui.adsabs.harvard.edu/abs/2019MNRAS.487.5572M} {487, 5572}

\bibitem[\protect\citeauthoryear{{Naab} et~al.,}{{Naab} et~al.}{2014}]{Naab+14}
{Naab} T.,  et~al., 2014, \mn@doi [\mnras] {10.1093/mnras/stt1919}, \href
  {https://ui.adsabs.harvard.edu/abs/2014MNRAS.444.3357N} {444, 3357}

\bibitem[\protect\citeauthoryear{{Newville} et~al.,}{{Newville}
  et~al.}{2020}]{lmfit}
{Newville} M.,  et~al., 2020, {lmfit/lmfit-py 1.0.1},
  \mn@doi{10.5281/zenodo.3814709}

\bibitem[\protect\citeauthoryear{{Oh} et~al.,}{{Oh} et~al.}{2020}]{Oh+20}
{Oh} S.,  et~al., 2020, \mn@doi [\mnras] {10.1093/mnras/staa1330}, \href
  {https://ui.adsabs.harvard.edu/abs/2020MNRAS.495.4638O} {495, 4638}

\bibitem[\protect\citeauthoryear{{Oke} \& {Gunn}}{{Oke} \&
  {Gunn}}{1983}]{Oke+83}
{Oke} J.~B.,  {Gunn} J.~E.,  1983, \mn@doi [\apj] {10.1086/160817}, \href
  {https://ui.adsabs.harvard.edu/abs/1983ApJ...266..713O} {266, 713}

\bibitem[\protect\citeauthoryear{{Owers} et~al.,}{{Owers}
  et~al.}{2017}]{Owers+17}
{Owers} M.~S.,  et~al., 2017, \mn@doi [\mnras] {10.1093/mnras/stx562}, \href
  {https://ui.adsabs.harvard.edu/abs/2017MNRAS.468.1824O} {468, 1824}

\bibitem[\protect\citeauthoryear{{Proctor}, {Forbes}  \& {Beasley}}{{Proctor}
  et~al.}{2004}]{Proctor+04}
{Proctor} R.~N.,  {Forbes} D.~A.,   {Beasley} M.~A.,  2004, \mn@doi [\mnras]
  {10.1111/j.1365-2966.2004.08415.x}, \href
  {https://ui.adsabs.harvard.edu/abs/2004MNRAS.355.1327P} {355, 1327}

\bibitem[\protect\citeauthoryear{{Rawlings} et~al.,}{{Rawlings}
  et~al.}{2020}]{Rawlings+20}
{Rawlings} A.,  et~al., 2020, \mn@doi [\mnras] {10.1093/mnras/stz2797}, \href
  {https://ui.adsabs.harvard.edu/abs/2020MNRAS.491..324R} {491, 324}

\bibitem[\protect\citeauthoryear{{S{\'a}nchez} et~al.,}{{S{\'a}nchez}
  et~al.}{2012}]{Sanchez+12}
{S{\'a}nchez} S.~F.,  et~al., 2012, \mn@doi [\aap]
  {10.1051/0004-6361/201117353}, \href
  {https://ui.adsabs.harvard.edu/abs/2012A&A...538A...8S} {538, A8}

\bibitem[\protect\citeauthoryear{{S{\'a}nchez} et~al.,}{{S{\'a}nchez}
  et~al.}{2021}]{Sanchez+21}
{S{\'a}nchez} S.~F.,  et~al., 2021, \mn@doi [\aap]
  {10.1051/0004-6361/202141225}, \href
  {https://ui.adsabs.harvard.edu/abs/2021A&A...652L..10S} {652, L10}

\bibitem[\protect\citeauthoryear{{Saunders} et~al.,}{{Saunders}
  et~al.}{2004}]{Saunders+04}
{Saunders} W.,  et~al., 2004, in {Moorwood} A. F.~M.,  {Iye} M.,  eds,  Society
  of Photo-Optical Instrumentation Engineers (SPIE) Conference Series Vol.
  5492, Ground-based Instrumentation for Astronomy. pp 389--400,
  \mn@doi{10.1117/12.550871}

\bibitem[\protect\citeauthoryear{{Scott} et~al.,}{{Scott}
  et~al.}{2017}]{Scott+17}
{Scott} N.,  et~al., 2017, \mn@doi [\mnras] {10.1093/mnras/stx2166}, \href
  {https://ui.adsabs.harvard.edu/abs/2017MNRAS.472.2833S} {472, 2833}

\bibitem[\protect\citeauthoryear{{Scott} et~al.,}{{Scott}
  et~al.}{2018}]{Scott+18}
{Scott} N.,  et~al., 2018, \mn@doi [\mnras] {10.1093/mnras/sty2355}, \href
  {https://ui.adsabs.harvard.edu/abs/2018MNRAS.481.2299S} {481, 2299}

\bibitem[\protect\citeauthoryear{{Segers}, {Schaye}, {Bower}, {Crain},
  {Schaller}  \& {Theuns}}{{Segers} et~al.}{2016}]{Segers+16}
{Segers} M.~C.,  {Schaye} J.,  {Bower} R.~G.,  {Crain} R.~A.,  {Schaller} M.,
  {Theuns} T.,  2016, \mn@doi [\mnras] {10.1093/mnrasl/slw111}, \href
  {https://ui.adsabs.harvard.edu/abs/2016MNRAS.461L.102S} {461, L102}

\bibitem[\protect\citeauthoryear{{Shanks} et~al.,}{{Shanks}
  et~al.}{2013}]{Shanks+13}
{Shanks} T.,  et~al., 2013, The Messenger, \href
  {http://adsabs.harvard.edu/abs/2013Msngr.154...38S} {154, 38}

\bibitem[\protect\citeauthoryear{{Shanks} et~al.,}{{Shanks}
  et~al.}{2015}]{Shanks+15}
{Shanks} T.,  et~al., 2015, \mn@doi [\mnras] {10.1093/mnras/stv1130}, \href
  {http://adsabs.harvard.edu/abs/2015MNRAS.451.4238S} {451, 4238}

\bibitem[\protect\citeauthoryear{{Sharp} et~al.,}{{Sharp}
  et~al.}{2006}]{Sharp+06}
{Sharp} R.,  et~al., 2006, in Society of Photo-Optical Instrumentation
  Engineers (SPIE) Conference Series. p. 62690G (\mn@eprint {}
  {astro-ph/0606137}), \mn@doi{10.1117/12.671022}

\bibitem[\protect\citeauthoryear{{Slipher}}{{Slipher}}{1914}]{Slipher+14}
{Slipher} V.~M.,  1914, Lowell Observatory Bulletin, \href
  {https://ui.adsabs.harvard.edu/abs/1914LowOB...2...66S} {2, 66}

\bibitem[\protect\citeauthoryear{{Smethurst} et~al.,}{{Smethurst}
  et~al.}{2015}]{Smethurst+15}
{Smethurst} R.~J.,  et~al., 2015, \mn@doi [\mnras] {10.1093/mnras/stv161},
  \href {https://ui.adsabs.harvard.edu/abs/2015MNRAS.450..435S} {450, 435}

\bibitem[\protect\citeauthoryear{{Smith} et~al.,}{{Smith}
  et~al.}{2004}]{Smith+04}
{Smith} G.~A.,  et~al., 2004, in {Moorwood} A. F.~M.,  {Iye} M.,  eds,  Society
  of Photo-Optical Instrumentation Engineers (SPIE) Conference Series Vol.
  5492, Ground-based Instrumentation for Astronomy. pp 410--420,
  \mn@doi{10.1117/12.551013}

\bibitem[\protect\citeauthoryear{{Springel}, {Di Matteo}  \&
  {Hernquist}}{{Springel} et~al.}{2005}]{Springel+05}
{Springel} V.,  {Di Matteo} T.,   {Hernquist} L.,  2005, \mn@doi [\apjl]
  {10.1086/428772}, \href
  {https://ui.adsabs.harvard.edu/abs/2005ApJ...620L..79S} {620, L79}

\bibitem[\protect\citeauthoryear{{Taylor} et~al.,}{{Taylor}
  et~al.}{2011}]{Taylor+11}
{Taylor} E.~N.,  et~al., 2011, \mn@doi [\mnras]
  {10.1111/j.1365-2966.2011.19536.x}, \href
  {https://ui.adsabs.harvard.edu/abs/2011MNRAS.418.1587T} {418, 1587}

\bibitem[\protect\citeauthoryear{{Thomas}, {Maraston}, {Schawinski}, {Sarzi}
  \& {Silk}}{{Thomas} et~al.}{2010}]{Thomas+10}
{Thomas} D.,  {Maraston} C.,  {Schawinski} K.,  {Sarzi} M.,   {Silk} J.,  2010,
  \mn@doi [\mnras] {10.1111/j.1365-2966.2010.16427.x}, \href
  {https://ui.adsabs.harvard.edu/abs/2010MNRAS.404.1775T} {404, 1775}

\bibitem[\protect\citeauthoryear{{Trager}, {Worthey}, {Faber}, {Burstein}  \&
  {Gonz{\'a}lez}}{{Trager} et~al.}{1998}]{Trager+98}
{Trager} S.~C.,  {Worthey} G.,  {Faber} S.~M.,  {Burstein} D.,   {Gonz{\'a}lez}
  J.~J.,  1998, \mn@doi [\apjs] {10.1086/313099}, \href
  {https://ui.adsabs.harvard.edu/abs/1998ApJS..116....1T} {116, 1}

\bibitem[\protect\citeauthoryear{{Trager}, {Faber}, {Worthey}  \&
  {Gonz{\'a}lez}}{{Trager} et~al.}{2000}]{Trager+00}
{Trager} S.~C.,  {Faber} S.~M.,  {Worthey} G.,   {Gonz{\'a}lez} J.~J.,  2000,
  \mn@doi [\aj] {10.1086/301442}, \href
  {https://ui.adsabs.harvard.edu/abs/2000AJ....120..165T} {120, 165}

\bibitem[\protect\citeauthoryear{{Tremaine} et~al.,}{{Tremaine}
  et~al.}{2002}]{Tremaine+02}
{Tremaine} S.,  et~al., 2002, \mn@doi [\apj] {10.1086/341002}, \href
  {https://ui.adsabs.harvard.edu/abs/2002ApJ...574..740T} {574, 740}

\bibitem[\protect\citeauthoryear{{Watson} et~al.,}{{Watson}
  et~al.}{2022}]{Watson+21}
{Watson} P.~J.,  et~al., 2022, \mn@doi [\mnras] {10.1093/mnras/stab3477}, \href
  {https://ui.adsabs.harvard.edu/abs/2022MNRAS.510.1541W} {510, 1541}

\bibitem[\protect\citeauthoryear{{Worthey} \& {Ottaviani}}{{Worthey} \&
  {Ottaviani}}{1997}]{Worthey+97}
{Worthey} G.,  {Ottaviani} D.~L.,  1997, \mn@doi [\apjs] {10.1086/313021},
  \href {https://ui.adsabs.harvard.edu/abs/1997ApJS..111..377W} {111, 377}

\bibitem[\protect\citeauthoryear{{de La Rosa}, {La Barbera}, {Ferreras}  \& {de
  Carvalho}}{{de La Rosa} et~al.}{2011}]{de_la_Rosa+11}
{de La Rosa} I.~G.,  {La Barbera} F.,  {Ferreras} I.,   {de Carvalho} R.~R.,
  2011, \mn@doi [\mnras] {10.1111/j.1745-3933.2011.01146.x}, \href
  {https://ui.adsabs.harvard.edu/abs/2011MNRAS.418L..74D} {418, L74}

\bibitem[\protect\citeauthoryear{{de Zeeuw} et~al.,}{{de Zeeuw}
  et~al.}{2002}]{de_Zeeuw+02}
{de Zeeuw} P.~T.,  et~al., 2002, \mn@doi [\mnras]
  {10.1046/j.1365-8711.2002.05059.x}, \href
  {https://ui.adsabs.harvard.edu/abs/2002MNRAS.329..513D} {329, 513}

\bibitem[\protect\citeauthoryear{{van de Sande} et~al.,}{{van de Sande}
  et~al.}{2017a}]{van_de_Sande+17a}
{van de Sande} J.,  et~al., 2017a, \mn@doi [\mnras] {10.1093/mnras/stx1751},
  \href {https://ui.adsabs.harvard.edu/abs/2017MNRAS.472.1272V} {472, 1272}

\bibitem[\protect\citeauthoryear{{van de Sande} et~al.,}{{van de Sande}
  et~al.}{2017b}]{van_de_Sande+17b}
{van de Sande} J.,  et~al., 2017b, \mn@doi [\apj]
  {10.3847/1538-4357/835/1/104}, \href
  {https://ui.adsabs.harvard.edu/abs/2017ApJ...835..104V} {835, 104}

\bibitem[\protect\citeauthoryear{{van de Sande} et~al.,}{{van de Sande}
  et~al.}{2018}]{van_de_Sande+18}
{van de Sande} J.,  et~al., 2018, \mn@doi [Nature Astronomy]
  {10.1038/s41550-018-0436-x}, \href
  {https://ui.adsabs.harvard.edu/abs/2018NatAs...2..483V} {2, 483}

\bibitem[\protect\citeauthoryear{{van de Sande} et~al.,}{{van de Sande}
  et~al.}{2021a}]{van_de_Sande+21a}
{van de Sande} J.,  et~al., 2021a, \mn@doi [\mnras] {10.1093/mnras/stab1490},
  \href {https://ui.adsabs.harvard.edu/abs/2021MNRAS.505.3078V} {505, 3078}

\bibitem[\protect\citeauthoryear{{van de Sande} et~al.,}{{van de Sande}
  et~al.}{2021b}]{van_de_Sande+21b}
{van de Sande} J.,  et~al., 2021b, \mn@doi [\mnras] {10.1093/mnras/stab2647},
  \href {https://ui.adsabs.harvard.edu/abs/2021MNRAS.508.2307V} {508, 2307}

\bibitem[\protect\citeauthoryear{{van den Bergh}}{{van den
  Bergh}}{1976}]{van_den_Bergh+76}
{van den Bergh} S.,  1976, \mn@doi [\apj] {10.1086/154452}, \href
  {https://ui.adsabs.harvard.edu/abs/1976ApJ...206..883V} {206, 883}

\makeatother
\end{thebibliography}

% Don't change these lines
\bsp	% typesetting comment
\label{lastpage}
\end{document}